\documentclass[titlepage]{article}
\usepackage{graphicx}
\usepackage[margin=1in]{geometry}
\usepackage{amssymb,amsfonts,amsmath,color}
\usepackage{booktabs}
\usepackage{cite}
\usepackage[mathscr]{euscript}
\usepackage[T1]{fontenc}
\usepackage{lmodern}
\usepackage[applemac]{inputenc}
\usepackage{tabu}
\usepackage{hyperref}
\usepackage{authblk}
\usepackage{sidecap}

\begin{document}
\setcounter{secnumdepth}{0}

\author{Christopher McFarland\thanks{Graduate Program in Biophysics, Harvard University, Boston, MA 02115} 
\and 
Leonid Mirny\textsuperscript{1, }\thanks{Department of Physics, Massachusetts Institute of Technology (MIT), Cambridge, MA 02139, USA}\textsuperscript{,}
\thanks{Institute for Medical Engineering and Science, MIT, Cambridge, MA 02139, USA}\textsuperscript{, *} 
\and 
Kirill S. Korolev\textsuperscript{2, }\thanks{Currently at Department of Physics, Boston University, Boston, MA 02215
\newline \indent $\ $*Corresponding Author: \href{leonid@mit.edu}{lenoid@mit.edu}, E25-526, MIT, 77 Massachusetts ave, Cambridge, MA 02139, (617) \newline \indent$\ $452-4862}}

\title{A tug-of-war between driver and passenger mutations in cancer and other adaptive processes}



\maketitle

\newcommand{\Pcancer}[0] {
  P_{\rm{cancer}}}

\newcommand{\Tcancer}[0] {
  t_{\rm{cancer}}}


\begin{abstract}
Cancer progression is an example of a rapid adaptive process where evolving new traits is essential for survival and requires a high mutation rate. Precancerous cells acquire a few key mutations that drive rapid population growth and carcinogenesis. Cancer genomics demonstrates that these few `driver' mutations occur alongside thousands of random `passenger' mutations---a natural consequence of cancer's elevated mutation rate. Some passengers can be deleterious to cancer cells, yet have been largely ignored in cancer research. In population genetics, however, the accumulation of mildly deleterious mutations has been shown to cause population meltdown. Here we develop a stochastic population model where beneficial drivers engage in a tug-of-war with frequent mildly deleterious passengers. These passengers present a barrier to cancer progression that is described by a critical population size, below which most lesions fail to progress, and a critical mutation rate, above which cancers meltdown. We find support for the model in cancer age-incidence and cancer genomics data that also allow us to estimate the fitness advantage of drivers and fitness costs of passengers. We identify two regimes of adaptive evolutionary dynamics and use these regimes to rationalize successes and failures of different treatment strategies. We find that a tumor's load of deleterious passengers can explain previously paradoxical treatment outcomes and suggest that it could potentially serve as a biomarker of response to mutagenic therapies. Collective deleterious effect of passengers is currently an unexploited therapeutic target. We discuss how their effects might be exacerbated by both current and future therapies.
\end{abstract}

\section{Introduction}
While many populations evolve new traits via a gradual accumulation of changes, some adapt very rapidly. Examples include viral adaptation during infection (1); the emergence of antibiotic resistance (2); artificial selection in biotechnology (3); and cancer (4).  Rapid adaptation is characterize by three key features: (i) the availability of strongly advantageous traits accessible by a few mutations, (ii) an elevated mutation rate (5, 6), and (iii) a dynamic population size (7). Traditional theories of gradual adaptation are not applicable under these conditions, and new approaches are needed to address pressing problems in medicine and biotechnology.

Cancer progression is an example of a rapidly adapting population: cancers develop as many as ten new traits (8), often have a high mutation rate (8-10), and a population size that is rapidly changing over time. This process is driven by a handful of mutations and chromosomal abnormalities in cancer-related genes (oncogenes and tumor suppressors), collectively called \emph{drivers}. From an evolutionary point of view, drivers are mutations that are beneficial to cancer cells because their phenotypes increase the cell proliferation rate or eliminate brakes on proliferation (8). Drivers, however, arise alongside thousands of other mutations and alterations dispersed through the genome that have no immediate beneficial effect, collectively called \emph{passengers}. 

Passengers have been previously assumed to be neutral and largely ignored in cancer research, yet growing evidence suggests that they may sometimes be deleterious to cancer cells and, thus, play an important role in both neoplastic progression and clinical outcomes. In an earlier study, we showed that deleterious passengers can readily accumulate during tumor progression and found that many passengers present in cancer genomes exhibit signatures of damaging mutations (11). Additionally, chromosomal gains and losses that are pervasive in cancer can be passengers, and have been shown to be highly damaging to cancer cells (12). Lastly, cancers with high levels of chromosomal alterations exhibit better clinical outcomes in breast, ovarian, gastric, and non-small cell lung cancer (13). Passenger mutations and chromosomal abnormalities can be deleterious via a variety of mechanisms: direct loss-of-function (14), proteotoxic cytotoxicity from protein disbalance and aggregation (15), or by inciting an immune response (16).

While the role of deleterious mutations in cancer is largely unknown, their effects on natural populations has been extensively studied in genetics (5, 17-19). The accumulations of deleterious mutations can cause the extinction of a population via processes known as Muller's ratchet and mutational meltdown (17, 20, 21) It was recently proposed that inevitable accumulation of deleterious mutations in natural populations should be offset by new beneficial mutations, leading to long-term population stability (19). Here we consider a rapid adaptation of a population with a variable size and subject of a high mutation rate. A rapidly adapting population faces a double bind: it must quickly acquire, often exceeding rare, adaptive mutations and yet avoid mutational meltdown. As a result, adaptive processes frequently fail. Indeed, less than 0.1\% of species on earth have adapted fast enough to avoid extinction (22) and, similarly, only about 0.1\% of precancerous lesions ever advance to cancer (23). To control cancer or pathogens, we should understand the constraints that evolution imposes on their rapid adaptation. 

Here we investigate how asexual populations such as tumors rapidly evolve new traits while avoiding mutational meltdown. Unlike classical theories of gradual adaptation, the evolutionary model we develop has three key features: (i) rare, strongly advantageous driver mutations, (ii) a high mutation rate that makes moderately deleterious passengers relevant, and (iii) a population size that varies with the fitness of individual cells. We found that a tug-of-war between beneficial drivers and deleterious passengers creates two major regimes of population dynamics: an adaptive regime, where the probability of adaptation (cancer) is high; and a non-adaptive regime, where adaptation (cancer) is exceedingly rare. 

Adaptive and non-adaptive regimes are separated by a critical population size or barrier to cancer progression that most lesions fail to overcome, and a critical mutation rate that leads to mutational meltdown. We found strong evidence of these phenomena in age-incidence curves and recent cancer genomics data. Agreement of the model with these data allows us to estimate the selective advantages of drivers as 10-50\%, a range consistent with recent direct experimental measurements (24). Genomic data also show that deleterious passengers are approximately 100 times weaker. Our model offers a new interpretation of cancer treatment strategies and explains a previously paradoxical relationship between cancer mutation rates and therapeutic outcomes. Most importantly, it suggests that deleterious passengers offer a new, unexploited avenue of cancer therapy. 

\section{Results}
\subsection{Model}
We consider a dynamic population of cells that can divide, mutate (in a general sense, i.e. including alterations, epigenetic changes, etc), and die stochastically. Mutations occur during cell divisions with a per-locus rate $\mu$. The number of \textbf{d}river loci in the genome, i.e. a driver target size, is $T_d$, while the target size for deleterious \textbf{p}assengers is $T_p$. Hence, the genome-wide driver and passenger mutation rates are $\mu_d = \mu T_d$ and $\mu_p = \mu T_p$ respectively. A driver increases an individual's growth rate by $s_d \sim 0.1$, while a new passenger decreases the growth rate by $s_p \sim 10^{-4}-10^{-1}$. Here we only consider fixed values of $s_d$ and $s_p$ because previous work showed that drivers and passengers sampled from various fitness distributions (exponential, normal, and Gamma) exhibit essentially the same dynamics (11). The net effect of multiple mutations on cell fitness w is given by $w={{\left( 1+{{s}_{d}} \right)}^{{{n}_{d}}}}{{\left( 1+{{s}_{p}} \right)}^{{{n}_{p}}}}$, where $n_d$ and $n_p$ are the total number of drivers and passengers in a cell. 

The birth and death rates of a cell in our model depend not only on fitness, but also on the population size $N$ via a Gompertzian growth function often used to describe cancerous populations (25) (see \textbf{SI} for details). At large $N$, deaths exceed births and tumors must adapt (or innovate) via new drivers to progress to larger population sizes. Thus, populations in our model expand and shrink in two ways: on a short time-scale due to stochastic cell divisions and deaths, and on a long time-scale due to the accumulation of advantageous and deleterious mutations. Previous models of advantageous and deleterious mutations have not considered a varying population size (26, 27).
 
In cancer and other adapting populations the target size for advantageous mutations (drivers) is much smaller than the target size for deleterious mutations $(T_d \ll T_p)$. If driver loci include a few specific sites ($\sim 10$ per gene) in all cancer-associated genes (approximately 100, (28)), then collectively drivers will constitute less than one one-millionth of the genome. Conversely, as much as 10\% of the human genome is well-conserved and likely deleterious when mutated (29, 30). In natural populations, $T_p$ should still remain much greater than $T_d$ simply because natural selection optimizes genomes to their environment, implying that most changes will be neutral or damaging. Indeed, most protein coding mutations and alterations were deleterious or neutral when investigated in fly (31), yeast (32), and bacterial genomes (33). We consider only moderately deleterious loci here $(s_p \approx 10^{-4} - 10^{-1})$---which account for most nonsynonymous mutations (34, 35). Deleterious mutations outside of this range either do not fixate or negligibly alter progression (11). Hence, we used a conservative size of $T_p \approx 10^{5} - 10^{7}$ loci to account for passengers with fitness effects outside of this range that we are neglecting (see \textbf{SI} and \textbf{Table S1} for details of parameters estimation). This quantity is still much greater than $T_d$. Finally, we explored a variety of driver fitness advantages, as estimates in the literature ranged from 0.0001 (36) to 1 (24).

\subsection{A critical population size}
\textbf{Figure 1A} shows the dynamics $N(t)$ of individual populations starting at different initial sizes $N^0$, which correspond to different potential hyperplasia sizes (we begin trajectories immediately after a stem cell acquires its first driver, see \textbf{SI} for a discussion of dynamics before this time point). Populations exhibit two ultimate outcomes: growth to macroscopic size (i.e. cancer progression), or extinction. The prevalence of either outcome is determined by a critical population size $N^*$, about which larger populations $(N>N^*)$ generally commit to rapid growth and smaller populations $(N<N^*)$ generally commit to extinction. 

To understand the cause of this critical population size $N^*$, we looked at the short-term dynamics of populations. All trajectories show a reversed saw-toothed pattern (\textbf{Fig. 1B}), which result from a tug-of-war between drivers and passengers (11). When a new driver arises and takes over the population, the population size increases to a new stationary value. In between these rare driver events, the population size gradually decreases due to the accumulation of deleterious passengers. The relative rate of these competing processes determines whether a population commits to rapid growth or goes extinct. 

We can identify the location of $N^*$ by considering the average change in population size over time $(<dN/dt>)$, which is simply the average population growth due to driver accumulation $(v_d)$ minus the population decline due to passenger accumulation $(v_p)$. Fixation of a new driver causes an immediate jump in population size $\Delta N = N s_d$. These jumps occur randomly at a nearly constant rate $f = \mu_d N s_d$, given by the driver occurrence rate $\mu_d N$, multiplied by a driver's fixation probability $s_d/(1 + s_d) \approx s_d$. Hence, the velocity due to drivers is $v_d = f \Delta N = \mu_d N^2 s_d^2$. Similarly, passengers' velocity $(v_p = \mu_p N s_p)$ is a product of their rate of occurrence $(\mu_p N)$; their effect on population size $(N s_p)$; and their probability of fixation ($\sim 1/N$, a more accurate measure of this probability is used below and provided in the \textbf{SI}). Thus, we obtain:
\begin{equation}
\left\langle \frac{dN}{dt} \right\rangle ={{\mu }_{p}}{{s}_{p}}N\left( \frac{N}{{{N}^{*}}}-1 \right)
\end{equation}
\begin{equation}
{{N}^{*}}=\frac{{{T}_{p}}{{s}_{p}}}{{{T}_{d}}s_{d}^{2}}
\end{equation}
where $N^*$ is the critical population size. 

Because the population velocity is negative below $N^*$ and positive above $N^*$ there is an effective barrier for cancer (\textbf{Fig. 1C}): smaller populations tend to shrink, while larger populations tend to expand. Simulations support our conclusion that the probability of cancer increases with $N^0$ and sharply transitions at $N^*$ (\textbf{Fig. 1D}). Indeed, drastically different probability curves collapse onto a single curve once $N^0$ is rescaled by $N^*$ (computed from equation 2).  Since $N^*$ captures only the average, or mean-field, dynamics, it misses the variability of outcomes in rapidly adapting populations. \textbf{Figure 1E} illustrates that the variability of outcomes depends upon the strength of drivers $s_d$. Higher values of $s_d$ lead to larger stochastic jumps, which leads to larger deviations from mean behavior and more gradual changes in the probability of cancer across $N^0$. Thus, we formulated and analytically solved a stochastic generalization of equation 1 that incorporates this variability (\textbf{SI}). Our solution provides an excellent fit to simulations (\textbf{Fig. 1E}) and indicates that $N^*$ and $s_d$ fully describe population dynamics (\textbf{SI}). 

We can understand how $N^*$ and $s_d$ control cancer progression using a simple random-walk analogy. The population size experiences random jumps, resulting from driver fixation events, which are described by equation 1. These random jumps and declines are effectively a random walk in a one-dimensional effective potential (${{U}_{eff}}=\int{\left( {dN}/{dt}\; \right)dN}$), \textbf{Fig. 1C} and \textbf{SI}) with stochastic jumps of frequency $f$ and size $\Delta N$. Similar to chemical reactions activated by thermal energy, cancer progression is a rare event triggered by a quick succession of driver fixations. Below, we show that human tissues operate in a regime where progression is rare and successful lesions are the rare lesions that happen to acquire drivers faster than average. 
We found that population dynamics depend entirely on two dimensionless parameters: a deterministic mean velocity, dependent only upon $N/N^*$, and a stochastic step-size that is approximately proportional to $s_d$. By reducing the complexity of our evolutionary system to two parameters, we were next able to infer their values for real cancers without over-fitting. 

\begin{figure}[ht]
\begin{center}
\centerline{\includegraphics[width=\textwidth]{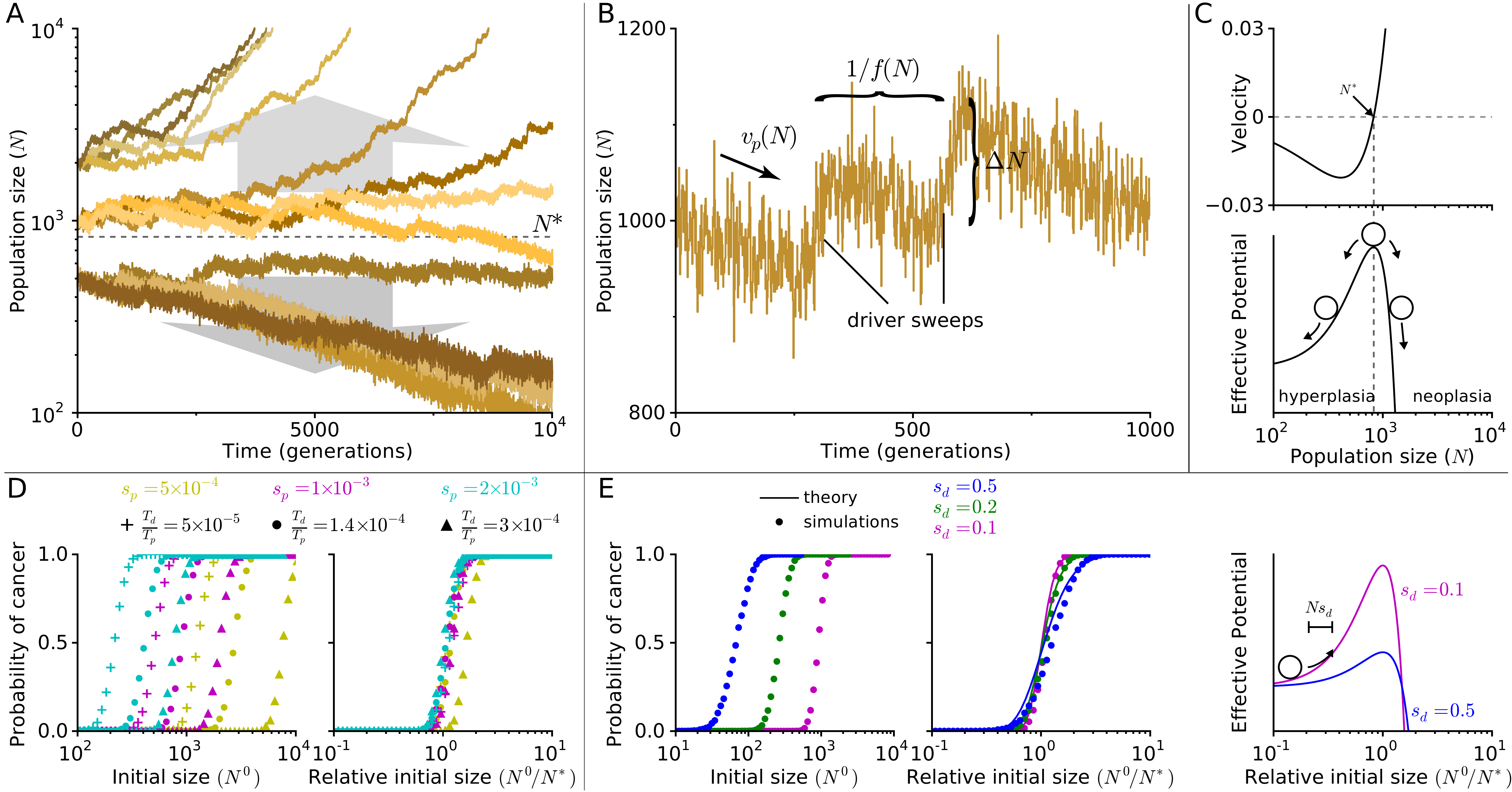}}
\caption{\textbf{Tug-of-war between drivers and passengers leads to a critical population size}
(A) Population size verses time of simulations initiated at various sizes. Populations starting above $N^*$ generally commit to rapid growth (i.e. adaptation) and to extinction below it. 
(B) A fragment of a trajectory shows periods of rapid growth and gradual decline. New drivers arrive with a frequency $f(N)$ and abruptly increases the population size by an amount $\Delta N$. Meanwhile, passenger accumulation causes populations to gradually decline with rate $v_p$. 
(C) Analytically computed mean velocity of population growth (top) and an effective barrier (bottom) as a function of population size $N$. The velocity is negative below $N^*$ and positive above it. 
(D) The probability of adaptation (cancer) as a function of initial population size $N$ (left) and a relative initial population size ($N/N^*$, right) for nine values of evolutionary parameters. Using the relative size $N/N^*$ leads to curve collapse, whereby populations with different evolutionary parameters nevertheless behave similarly. 
(E) Same as in (D) for simulations and theory but for different values of $s_d$. Higher values of $s_d$, leads to more gradual transition from non-adaptive to adaptive regime. Excellent agreement between simulations and theory demonstrates accuracy of the theory. In our formalism, an increase in $s_d$ results in a larger jump size $\Delta N$ and lower potential barrier, allowing more populations to overcome the barrier (right).}
\end{center}
\end{figure}

\subsection{Model validation using cancer incidence and genomic data}
Our model of cancer progression predicts the presence of an effective barrier to cancer where small lesions are very unlikely to ever progress to cancer. It also predicts a specific distribution in the number of passenger mutations and a specific relationship between drivers and passengers in individual cancer samples. We looked for evidences of these phenomena in age-incidence data (37) and cancer genomics data (28, 38-40). These comparisons also allowed us to estimate some critical parameters of the model: $N^0$, $s_d$, and $s_p$. 

\textbf{Figure 2A} presents the incidence rate of breast cancer versus age (37) alongside the predictions from a classic driver-only model (\textbf{SI}) and our model. The incidence rate was calculated by considering a process where precancerous lesions arise with a constant rate $r$ beginning at birth. These lesions then progress to cancer in time $\tau$ with a probability $P(\tau)$ that we determined from simulations (\textbf{Fig. S1}). By convoluting this distribution $P(\tau)$ with the lesion initiation rate $r$, we can predict the age-incidence rate $I(t)$. Because many lesions go extinct in our model and never progress to cancer, the predicted incidence rate saturates at old-age: ${{I}_{\max }}=r\int\limits_{0}^{\infty }{P\left( \tau  \right)d\tau =}r{{P}_{\infty }}$, where $P_{\infty}$ is the probability that a lesion will ever progresses to cancer, determined above. 

Both the observed population incidence rates and our driver-passenger model saturate with age. This is a direct result of the probability of progression from a lesion to cancer being low. We estimate a lower bound for the rate of lesion formation $r$ in breast cancer of at least 10 lesions per year that can be arrived at through two separate considerations: first, by considering the quantity of breast epithelial stem cells and the rate at which they can mutate into lesions (\textbf{SI}), and second, by considering the number of lesions observed within the breast tissue of normal cadavers (23). By comparing this limit ($\sim10\ \rm{lesions} \cdot \rm{year}^{-1}$) to the maximum observed breast cancer incidence rate $I_{\rm{max}} \approx 10^{-2}\ \rm{cancers} \cdot year^{-1}$, we find that $P_{\infty} \approx 10^{-3}$, or only about 1 in 1,000 lesions ever progress. This finding is consistent with a number of clinical studies that have observed that very few lesions ever progress to cancer, while many more regress to undetectable size (41, 42)---another property seen in our model. Thus, good agreement between age-incidence data and our model is obtained when $s_d \approx 0.1-0.2$ and $N^0/N^*$ is chosen such that $P_{\infty} = 10^{-3}$. This suggests that cancer begins at a population size far below $N^*$, where drivers are most often overpowered by passengers. Indeed, 21 of the 25 most prevalent cancers plateau at old-age suggesting that progression is inefficient in most tumor types (\textbf{Fig. S1}). In a driver-only model (see \textbf{SI} for details), every lesion progresses to cancer after sufficient time (i.e. $P_{\infty} = 1$), therefore a plateau in incidence rate can only result from a very low lesion formation rate (0.01 per year), which is inconsistent with abundant pathology data (23, 43). 

Recent cancer genomics data offer a new opportunity to validate our model. Specifically, we looked at Somatic Nonsynonymous Mutations (SNMs) and Somatic Copy-Number Alterations (SCNAs) derived from over 700 individual cancer-normal sample pairs obtain from the breast (38), colon (28), lung (40), and skin (39) (\textbf{Table S2}). We found similar results when analyzing SNMs and SCNAs both separately (\textbf{Fig. S2}, \textbf{Table S3}) or in aggregation (\textbf{Fig. 2BC}). \textbf{Figure 2B} shows a wide and asymmetric distribution of the total number of mutations, which is consistent with our model under realistic parameters.  A driver-only model yields a narrower and more symmetric distribution that is inconsistent with the data (\textbf{Fig. 2B}). The driver-only model can fit the observed distribution only if it assumes that just 1-2 drivers are needed for cancer (\textbf{Fig. S3}, \textbf{Table S4})---a value inconsistent with both the extent of recurrent mutations seen in cancer (10) and known biology. This large variance in mutation totals further supports our model and suggests that driver mutations and alterations have a large effect size: $s_d \approx 0.4$. Our estimate of $s_d \approx 0.1-0.4$ obtained from age-incidence and mutation histograms is in excellent agreement with experimentally measured changes in the growth rate of mouse intestinal stem cells upon induction of p53, APC or k-RAS mutations where measured values ranged from of 0.16 to 0.58 (24).
 
We then used cancer genomics data to compare the number of drivers and passengers observed in individual cancer samples to our model's predicted relationship. In our model, additional passengers must be counterbalanced by additional drivers for the population to succeed. If a lesion lingers around $N^*$ for a long time, then it must have acquired both many passengers and many counterbalancing drivers; while lesions that quickly progress through the barrier at $N^*$ acquire fewer of each. As a result, we expect a positive linear relationship between the number of drivers and passengers: $n_d \cdot s_d - n_p \cdot s_p = \rm{constant}$; this result follows directly from the definition of fitness in our model (\textbf{SI}). Our predicted positive linear relationship between drivers and passengers is indeed observed in all tumor types that we studied (\textbf{Fig. 2C}, \textbf{Table S3}, $p < 0.08-10^{-6}$). The slope of this regression line is predicted to be $s_p/s_d$, which ranged from 1/21 to 1/193 (\textbf{Table S3}) for the various subtypes. While there is considerable variation and large margins of error in these numbers, these slopes ($s_p/s_d  \approx  5 \cdot (10^{-2} - 10^{-3})$) correspond to an $s_p$ of $5 \cdot (10^{-3} - 10^{-4})$ when $s_d = 0.1$. These rough values are similar to germ-line SNMs in humans of European descent, where 64\% of all mutations exhibit an $s_p$ between $10^{-5}$ and $10^{-2}$ (35).
 
We considered and refuted several alternative explanations for the observed positive linear relationship between drivers and passengers. First, that the strength of SCNAs may differ from SNMs. Hence, we investigated each alteration-type separately and found positive linear relationships in both cases (\textbf{Fig. S2}, \textbf{Table S3}). Second, that the number of driver alterations might be explained by variation in the tumor stage, or the rate and/or mechanism of mutagenesis. In \textbf{Table S3} we show that these factors cannot suppress the correlation between drivers and passengers. Lastly, we considered and refuted the possibility that this relationship between drivers and passengers is non-linear (\textbf{Fig. 2C} insert). Because the data disagrees with all of these alternate hypotheses, we believe that it supports our conclusion that cancer progression is a tug-of-war between drivers and passengers.

\begin{SCfigure} 
\centering
\includegraphics[width=0.5\textwidth]{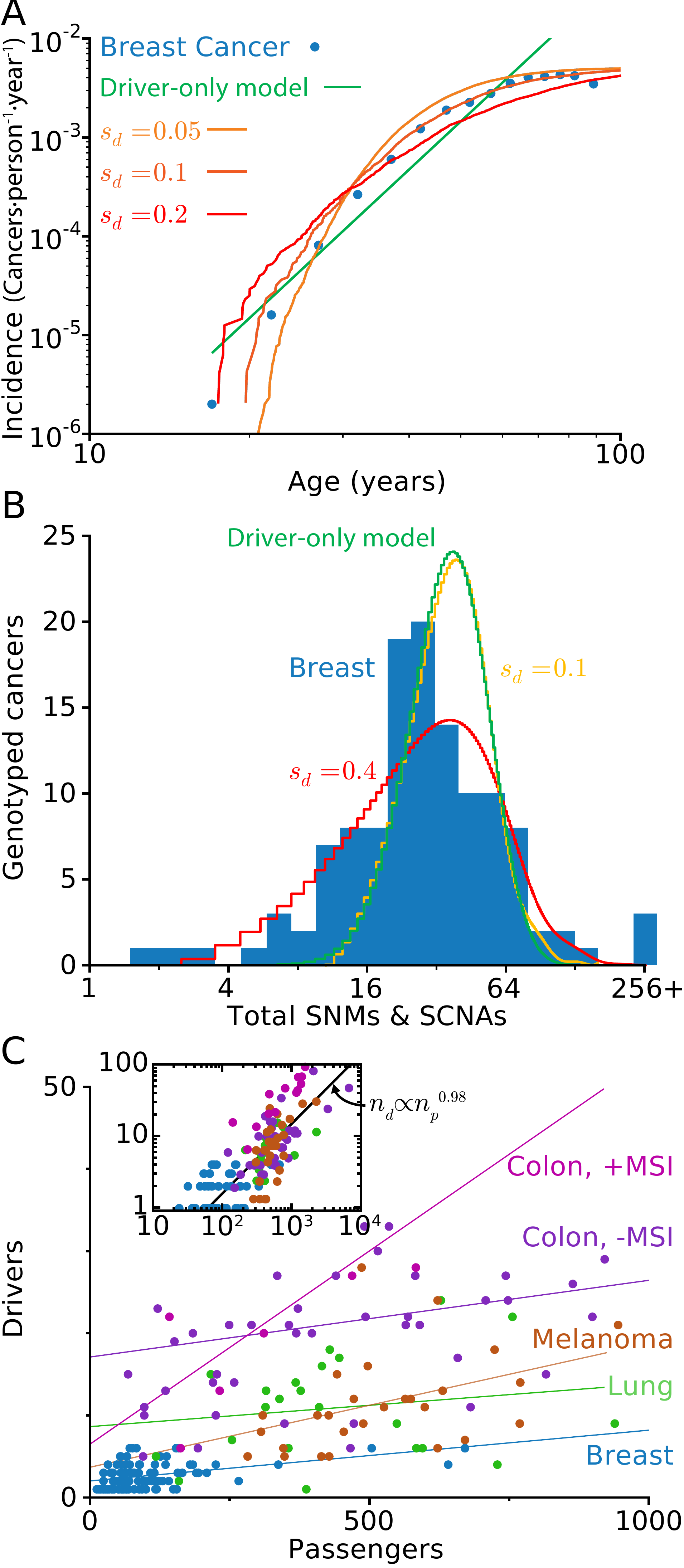}
\caption{\textbf{Signatures of balance between drivers and passengers in incidence and genomics data}
(A) Predicted and observed breast cancer incidence rates verses age. Notice that in our model, as well as in the data, incidence rates plateau at old age. A traditional driver-only model, where the incidence rate $I$ increases with patient age $t$ according to a power-law ($I \propto t^k$), does not saturate. 
(B) Histogram of the collective number of protein-coding mutations (SNMs) and alterations (SCNAs) in breast cancer alongside predicted distributions. Our model, captures the width and asymmetry of the distribution well for $s_d = 0.4$, while a driver-only model predicts a narrower and symmetric distribution. 
(C) The total number of driver verses the number of passenger alterations in sequenced tumors for several major subtypes. SCNAs and SNMs were aggregated. As predicted by the model, all subtypes exhibited a positive linear relationship between the number of drivers and passengers ($p < 0.08 - 10^{-5}$). A driver-only model with neutral passengers does not predict this linear relationship between drivers and passengers. 
(Insert) The same genomics data plotted on log axes, with the y-intercept from each subtype's linear fit subtracted. A linear relationship on logarithmic axes with a slope of approximately one suggests that the relationship between drivers and passengers is indeed linear.}
\end{SCfigure}

\subsection{A critical mutation rate}
We next used simulations to investigate the probability of cancer over a broad range of evolutionary parameters (\textbf{Fig. S4}) and found that there is a critical mutation rate above which the probability of cancer is exceedingly low (\textbf{Fig. 3A}). To explain this phenomenon and to find the parameters that determine this critical mutation rate $\mu^*$, we modified our analytical framework to consider selection against passengers and the effects of unfixed passengers on the accumulation of drivers. The modified framework, described in the SI, explains observed dynamics well (\textbf{Fig. 3AB}, \textbf{S4}). Previous theoretical work has shown that the number of unfixed passengers per cell is Poisson distributed with mean $\mu_p / s_p$ [first described in (44)]. This result assumes an approximate balance between the mutation rate of passengers and the selection against them, otherwise known as mutation-selection balance. The average fitness reduction of a cell due to this mutational load (i.e. the reduction in fitness relative to the fittest cells in the population) is $\mu_p$. A new driver arises in one of these cells at random and must carry the load of passengers residing in its cell along with it to fixation (18) (\textbf{Fig. 3C}); this process is often referred to as hitchhiking, so we describe these passengers as `hitchhikers'. 
If the reduction in fitness due to the load of passengers ($\mu_p = \mu T_p$) exceeds the benefit of a new driver ($s_d$), then the driver will not fixate (\textbf{Fig. 3C}). Hence, cancer is extremely rare when $\mu_p > s_d$. This suggests a critical mutation rate:
\begin{equation}
\mu^* = s_d/T_p
\end{equation}
The critical mutation rate suggests a new mode by which mutational meltdown operates. Prior models of mutational meltdown consider deleterious mutations in isolation (17), whereas our model points at the ability of deleterious mutations to inhibit the accumulation of advantageous mutations as a mechanism of meltdown. While it has been previously shown that deleterious mutations interfere with the fixation of beneficial alleles (18, 26, 45), this phenomenon has never been studied in the context of population survival. We discuss some important implications of this critical mutation rate for cancer treatment below.

We found support for the critical mutation rate in both cancer age-incidence and cancer genomics data. If we constraint cancer progression to develop within the typical timeframe for cancer progression (i.e. when we begin to see a plateau in incidence: $\sim 60$ years or 10,000 generations), the probability of cancer exhibits an optimum across mutation rates  (\textbf{Fig. 3D}). Above $\mu^*$ population meltdown is very common, while at very low mutation rates progression is too slow. The optimal mutation rate ($10^{-9} - 10^{-8} \rm{mutations} \cdot \rm{nucleotide}^{-1} \cdot \rm{generation}^{-1}$) is similar to mutation rates observed in cancer cell lines with a mutator phenotype (9) and the inferred mutation rate derived from the median number of mutations observed in a pan-cancer study of >3,000 tumors (10). Because $\mu^*$ depends only on $s_d$ and $T_p$, and is independent of other variables, we believe the maximal mutation rate should be the same across tumor subtypes (\textbf{SI}, \textbf{Fig. S4}). Indeed, a maximum of approximately 100 somatic mutations per Mb (99.4th percentile) was observed in the pan-cancer study mentioned above, which corresponds to our theoretical estimate (if we assume that the most mutagenic cancers still require ~1,000 generations to progress).

\begin{figure}[ht]
\begin{center}
\centerline{\includegraphics[width=\textwidth]{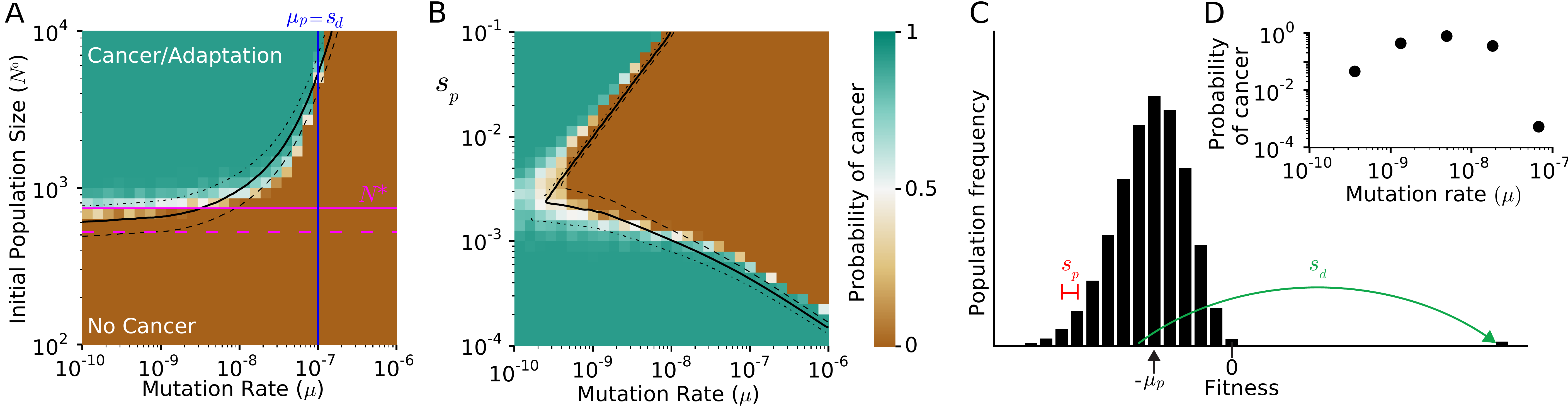}}
\caption{\textbf{Effect of mutation rate on cancer dynamics}
(A) The probability of cancer (adaptation) computed by simulations as a function of the initial population size and mutation rate. Evolutionary parameters roughly partition into a regime where cancer (adaption) is almost certain, and a regime where it is exceedingly rare. Estimates of $N^*$ from our theory (magenta, solid) accurately predict the transition observed at low mutation rates. Another transition is observed as mutation rate exceeds the critical mutation rate, also predicted by theory ($\mu^*$, blue line). Transition between these two regimes is better described by our theory when we incorporated (i) passenger interference with driver sweeps, and (ii) selection against passengers (black lines; dashed, solid, and dotted-dash predict 10\%, 50\%, and 90\% probability of adaptation). 
(B) Cancer (adaption) probability (color) obtained by simulations as a function of mutation rate and $s_p$. The theory (black lines) accurately reproduces the complex transition between both regimes. 
(C) Diagram illustrating how the load of passengers influences the probability of fixation of a driver. The distribution of fitness is due to a distribution of the number of passengers per cell, which follows a Poisson distribution with a mean reduction in fitness of $\mu_p$. Hence $s_p > \mu_p$ for a typical driver to outweigh the load of passengers. Segregating passengers not only reduce a driver's probability of fixation, but also its fitness benefit (26). 
(D) Probability of cancer for various mutation rates constrained to grow within 10,000 generations. We observe an optimum mutation rate.}
\end{center}
\end{figure}

\subsection{Two regimes of dynamics}
Taken together, our results demonstrate that the tug-of-war between advantageous drivers and deleterious passengers creates two major regimes of population dynamics: an adaptive regime where the probability of progression (cancer) is high ($\sim 100$ \%) and a non-adaptive regime where cancer progression is exceedingly rare (\textbf{Fig. 3}, \textbf{S4}). Evolving populations that fail to adapt and go extinct may do so because they reside in a non-adaptive region of the phase space. Similarly, normal tissues that avoid cancers may present a tumor microenvironment that is in this non-adaptive regime. By keeping $N^*$ sufficiently high, a tissue or clinician could keep cancerous populations outside of the adaptive regime. This critical population size $N^* = T_p s_p/(T_d s_d^2)$ depends on the evolutionary parameters of the system. For example, if $s_p$ were increased by tuning the response of the immune system to mutation-harboring cells, or if $T_d$ were decreased via a driver-targeted therapy, adaptation would become less likely. Below we demonstrate that a successful treatment must push a cancer back to the non-adaptive regime. 

\subsection{The adaptive barrier and critical mutation rate explain cancer treatment outcomes}
We simulated cancer growths and treatments and then monitored the long-term dynamics of these populations. Most treatments used today attempt to reduce tumor size, e.g. by specifically inhibiting key drivers (46) or by simply killing rapidly dividing cells (chemotherapy and radiation). Chemotherapy and radiation also elevate the mutation rate, thus affecting evolutionary dynamics. Previous work on the evolution of resistance to therapy has not considered the barriers to adaptation that we observe, so we re-investigated evolutionary outcomes from standard therapies and identified new potential ways to treat cancer. While real cancers have a varied evolutionary history, our analytical formalism predicts that cancer's future dynamics depend only on their current state, not their history (i.e. cancer dynamics are approximately path-independent, \textbf{Fig. S5}). We show below that this assumption can accurately predict outcomes in simulations and the clinic. 

In \textbf{Figure 4}, we present the evolutionary paths of cancer---from hyperplasia, to cancer, to treatment, and relapse or remission---on top of the phase diagrams described earlier. Our analysis demonstrates that a treatment is successful if it pushes a cancer into the non-adaptive regime of evolutionary dynamics where the probability of adaptation is low. Conversely, therapies fail, and populations re-adapt and remiss, when the therapy does not move cancer far enough to place it in the non-adaptive regime.

Our model suggests that chemotherapies succeed, in part, because they move cancers across the mutational threshold $\mu^*$. Beyond this threshold, the probability that a driver is strong enough to overpower a load of passengers becomes small (see above, \textbf{Fig. 2C}), making it hard for cancer to readapt. Increasing the mutation rate has little effect on the critical population size $N^*$ (see above). 

Thus, our model suggest that cancers with a very high load of mutations/alterations are close to the critical mutation rate and should be more susceptible to mutagenic treatments, such as chemotherapy. Several recent studies (13, 47) have noticed that patients survive breast and ovarian cancer most often when their tumors exhibited exceptional high levels of chromosomal alterations. This phenomenon was robust within and between subtypes of breast cancer (47). This finding is paradoxical for all previous models of cancer, where a greater mutation rate always accelerates cancer evolution and adaptation; yet is fully consistent with our model (\textbf{Fig. 4B}). 

Treatments exploiting the mutational load of cancers (i.e. their accumulated passengers) remain largely unexplored. We show that increasing the deleterious effect of passengers $s_p$ causes tumors to enter remission. Increasing $s_p$ is doubly effective because it exacerbates the deleteriousness of accumulated passengers and also slows down future adaptation. When we simulate such treatment by a 3-5 fold increase of $s_p$ (\textbf{Fig. 4C}), we observe an immediate decline in the population size followed by a low probability of replace due to an increased $N^*$. The phase diagram shows that a mild increase in $s_p$ is sufficient to push a population into an extinction regime and thus induce remission. Below we discuss possible treatment strategies that would increase $s_p$.

Given the large number of treatment options, finding therapies that work synergistically is a very important problem in cancer research (reviewed in (48)). While synergism is often discussed in the context of pharmacology, our phase diagrams identify \emph{evolutionarily} synergistic treatments. We found that remission was most likely to occur when the mutation rate and the fitness cost of passengers were increased simultaneously, more so than would be expected from simply adding together the effects of the individual therapies (\textbf{Fig. S6}). Hence, combinations of mutagenic chemotherapy along with treatments that elevate the cost of a mutational load may be most effective. According to our model, these therapies should also be compatible and complementary to driver-targeted therapies. 

\begin{figure}[ht]
\begin{center}
\centerline{\includegraphics[width=\textwidth]{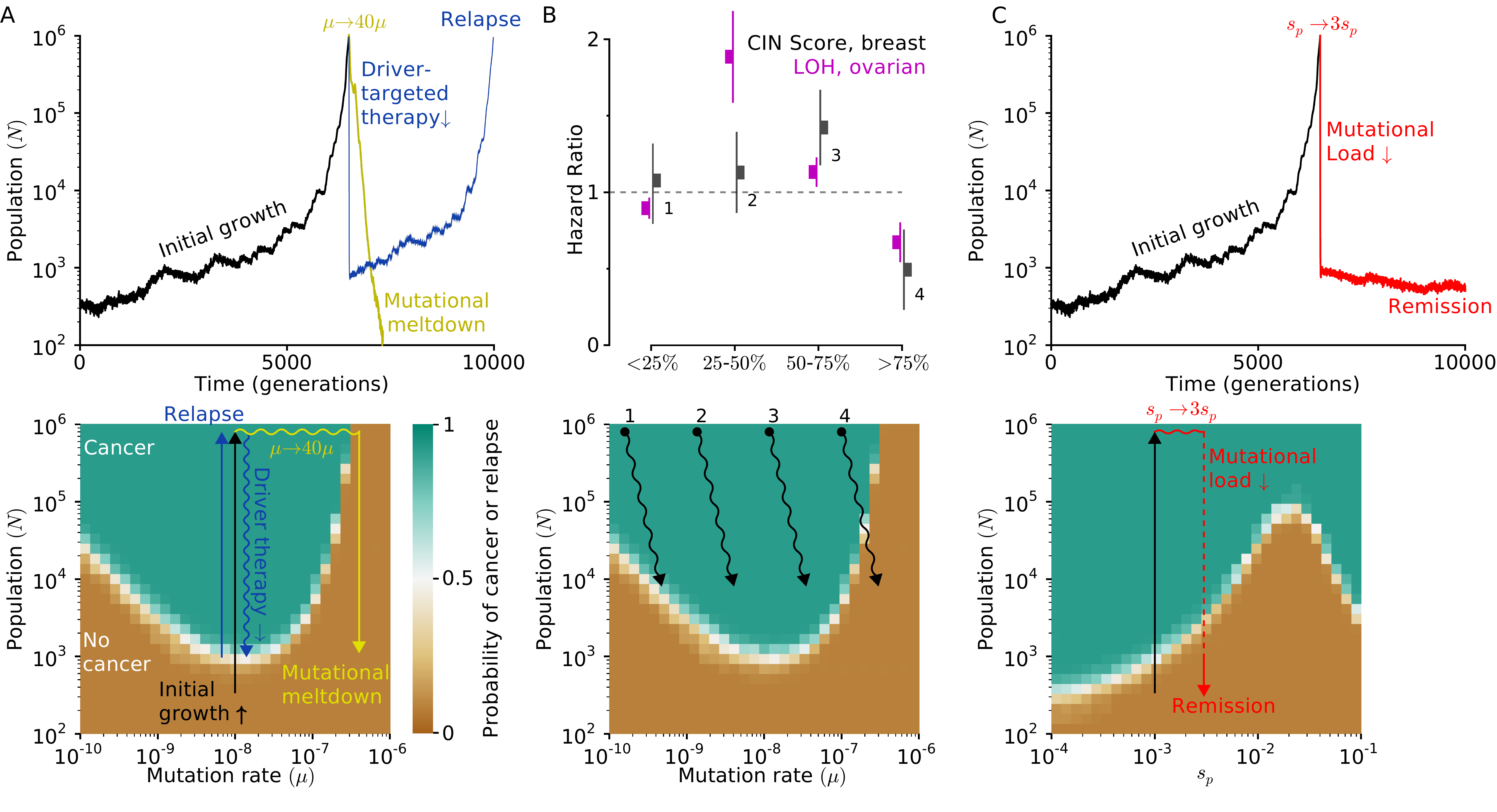}}
\caption{\textbf{Mapping and interpreting treatment outcomes}
(A) An adapted population (grown cancer) can be reverted to extinction by increasing the mutation rate (mutagenic chemotherapy) or by decreasing the population size (e.g. surgery or cytotoxic chemotherapy). Our phase diagrams explains therapeutic outcomes: therapies that reduce cancer size or increase mutation rate enough to push it outside of an adaptive regime cause continued population collapse; those that do not experience relapse. 
(B) (Top) Comparison of the model to clinical data. Cancers with intermediate mutational loads are the most aggressive (4, 13). Patients with intermediate Chromosomal INstability (CIN) and Loss Of Heterozygosity (LOH) scores are the least likely to survive. Patients with very high CIN are most effectively treated. (Bottom) Our phase diagrams shows these clinical outcomes: traditional therapies which work by decreasing the population size and/or increase the mutation rate work best for cancers with the highest mutation rate. 
(C) Three fold increase in the effect of passenger mutations leads to rapid population meltdown below $N^*$, thus relapse is unlikely.}
\end{center}
\end{figure}

\section{Discussion}
We present an evolutionary model of rapid adaptation that incorporates rare, strongly advantageous driver mutations and frequent, mildly deleterious passenger mutations. In this process, a population can either succeed and adapt, or fail and go extinct. We found theoretically, and confirmed by simulations, two regimes of dynamics: one where a population almost always adapts, and another where it almost never adapts. Complex stochastic dynamics, which emerge due to a tug-of-war between drivers and passengers, can be faithfully described as diffusion over a potential barrier that separates these two regimes. The potential barrier is located at a critical population size that a population must overcome to adapt. We also found a critical mutation rate, above which populations quickly meltdown. This general framework for adaptive asexual populations appears to be perfectly suited to characterize the dynamics of cancer progression and responses to therapy.

Progression to cancer is an adaptive process, driven by a few mutations in oncogenes and tumor suppressors. During this process, however, cells acquire tens of thousands of random mutations many of which may be deleterious to cancer cells. While strongly deleterious passenger mutations are weeded out by selection, mildly deleterious can fixate and even accumulate in a cancer by hitchhiking on drivers, as we have shown earlier (11). Passengers may be deleterious by inducing loss-of-function in critical proteins (14), gain-of-function toxicity via proteotoxic/misfolding stress (15, 49), or by triggering an immune response by a mutated epitope (16, 50). While we looked at passengers in cancer exomes and in SCNAs, passengers may also constitute epigenetic modifications, or karyotypic imbalances (15). Hence the number of deleterious passengers may be more than currently observed by genome-wide assays. 

Our framework suggests that most normal tissues reside in a regime where cancer progression is exceedingly rare; i.e. most lesions fail to grow above the critical population size and, thus, fail to overcome the adaptive barrier. Clinical cancers, on the contrary, reside above the adaptive barrier in a rapidly adapting state. Therapies must push a cancer below this adaptive barrier to succeed. In our framework, this entails moving the population below $N^*$ or increasing the mutation rate above $\mu^*$. The availability of a broad range of data for cancer allowed us to thoroughly test our framework's applicability.

We tested out model and estimated its parameters using cancer age-incidence curves, cancer exome sequences from almost 1,000 tumors in four cancer subtypes, and data on clinical outcomes. Age-incidence curves support the notion that the vast majority of lesions fail to progress and allow us to estimate the fitness benefit of a driver as $s_d \sim 0.1-0.4$. Genomics data suggests that passengers are indeed deleterious and that their deleterious effect is approximately one hundred times weaker than driver's beneficial effect. Moreover, the fitness benefit of a driver estimated from genomic data is roughly $s_d \sim 0.4$. Our range of values are consistent with recently measured 16-58\% increases in the mouse intestinal stem cell proliferation rate upon mutations in APC, k-RAS or p53 (24). Taken together these data support the notion of a tug-of-war between rare and large-effect drivers and frequent, but mildly deleterious passengers $s_p \sim 5 \cdot (10^{-4} - 10^{-3}$), which nevertheless have a large collective effect.

Results of our analysis have direct clinical implications.  Available clinical data (13, 47, 51) show that cancers with a higher load of chromosomal alterations, i.e. close to $\mu^*$, respond better to treatments. Our study suggests two potentially synergistic therapeutic strategies: to increase the mutation rate above $\mu^*$, and/or to increase the deleterious effect of accumulated passengers. An increase in the fitness cost of passengers would not only magnify the effects of an already accumulated mutational load, but also reduce future adaptation. This may be accomplished by (i) targeting unfolding protein response (UPR) pathways and/or the proteasome (15), (ii) hyperthermia that may further destabilize mutated proteins or clog UPR pathways (52), or (iii) by eliciting an immune response (16). Intriguingly, all these strategies are in clinical trials, yet they are often believed to work for reasons other than by exacerbating passengers' deleterious effects. In contrast, we predict that these therapies will be most effective in cancers with more passengers and an elevated mutation rate. Thus, characterizing the load of mutations/alterations in tumors may offer a new biomarker for predicting treatment outcomes and identify the best candidates for mutational chemotherapies.

While this study focused on asexual innovative evolution in cancer, our model may be generally applicable to other innovating populations. Consider a population in a new environment. The population is often initially small and fluctuating in size and often goes extinct, yet occasionally it expands to a larger stationary size by rapidly acquiring several new traits that are highly advantageous in the new environment. Both the evolutionary parameters (53) and observed phenomena (54) match our model well. Our mathematical framework may explain why these populations sometimes adapt, yet often fail.

\section{Materials and Methods}
All simulations were run using a previously-described first-order Gillespie algorithm (11). Driver genes were identified using MutSig (10) and GISTIC 2.0 (55) for potential NSM and SCNA drivers, respectively; requiring a Bonferroni-corrected enrichment $p$-value $\leq 5 \cdot 10^{-3}$ for classification of a gene as driver-harboring. All other genes were classified as `passenger-harboring'. 

\section{Acknowledgements}
CM and LM were supported by the National Cancer Institute-supported Physical Sciences in Oncology Center at MIT. KK was supported by Pappalardo Fellowship in Physics at MIT. We are grateful to Shamil Sunyaev, Gregory Krykov, Geoffrey Fudenberg, Maxim Imakaev, Anton Goloborodko for many productive discussions.

\section{References}
\begin{enumerate}
	\item Kawashima Y et al. (2009) Adaptation of HIV-1 to human leukocyte antigen class I. \emph{Nature} 458:641-5.
	\item Zhang Q et al. (2011) Acceleration of emergence of bacterial antibiotic resistance in connected microenvironments. \emph{Science} 333:1764-7.
	\item Wright SI et al. (2005) The effects of artificial selection on the maize genome. \emph{Science} 308:1310-4.
	\item Di Nicolantonio F et al. (2005) Cancer cell adaptation to chemotherapy. \emph{BMC Cancer} 5:78.
	\item Ishii K, Matsuda H, Iwasa Y, Sasaki A (1989) Evolutionarily stable mutation rate in a periodically changing environment. \emph{Genetics} 121:163-74.
	\item Moxon ER, Rainey PB, Nowak M a, Lenski RE (1994) Adaptive evolution of highly mutable loci in pathogenic bacteria. \emph{Curr Biol} 4:24-33.
	\item Cameron TC, O'Sullivan D, Reynolds A, Piertney SB, Benton TG (2013) Eco-evolutionary dynamics in response to selection on life-history. \emph{Ecol Lett} 16:754-63.
	\item Hanahan D, Weinberg R a (2011) Hallmarks of cancer: the next generation. \emph{Cell} 144:646-74.
	\item Jackson a L, Loeb L a (1998) The mutation rate and cancer. \emph{Genetics} 148:1483-90.
	\item Lawrence MS et al. (2013) Mutational heterogeneity in cancer and the search for new cancer-associated genes. \emph{Nature} 499:214-8.
	\item McFarland CD, Korolev KS, Kryukov G V, Sunyaev SR, Mirny L a (2013) Impact of deleterious passenger mutations on cancer progression. \emph{Proc Natl Acad Sci U S A} 110:2910-5.
	\item Williams BR, Amon A (2009) Aneuploidy: cancer's fatal flaw? \emph{Cancer Res} 69:5289-91.
	\item Birkbak NJ et al. (2011) Paradoxical relationship between chromosomal instability and survival outcome in cancer. \emph{Cancer Res} 71:3447-52.
	\item MacArthur DG et al. (2012) A systematic survey of loss-of-function variants in human protein-coding genes. \emph{Science} 335:823-8.
	\item Sheltzer JM, Amon A (2011) The aneuploidy paradox: costs and benefits of an incorrect karyotype. \emph{Trends Genet} 27:446-53.
	\item Segal NH et al. (2008) Epitope landscape in breast and colorectal cancer. \emph{Cancer Res} 68:889-92.
	\item Lynch M (2008) The cellular, developmental and population-genetic determinants of mutation-rate evolution. \emph{Genetics} 180:933-43.
	\item Johnson T, Barton NH (2002) The effect of deleterious alleles on adaptation in asexual populations. \emph{Genetics} 162:395-411.
	\item Good BH, Rouzine IM, Balick DJ, Hallatschek O, Desai MM (2012) Distribution of fixed beneficial mutations and the rate of adaptation in asexual populations. \emph{Proc Natl Acad Sci U S A} 109:4950-5.
	\item Zeyl C, Mizesko M, de Visser J a (2001) Mutational meltdown in laboratory yeast populations. \emph{Evolution} 55:909-17.
	\item Neher R a, Shraiman BI (2012) Fluctuations of fitness distributions and the rate of Muller's ratchet. \emph{Genetics} 191:1283-93.
	\item May, R. Lawton, J. Stork N (1995) \emph{Assessing Extinction Rates} (Oxford University Press).
	\item Wellings SR, Jensen HM, Marcum RG (1975) An atlas of subgross pathology of the human breast with special reference to possible precancerous lesions. \emph{J Natl Cancer Inst} 55:231-73.
	\item Vermeulen L et al. (2013) Defining stem cell dynamics in models of intestinal tumor initiation. \emph{Science} 342:995-8.
	\item Domingues JS (2012) Gompertz Model?: Resolution and Analysis for Tumors. \emph{J Math Model Appl} 1:70-77.
	\item Bachtrog D, Gordo I (2004) Adaptive evolution of asexual populations under Muller's ratchet. \emph{Evolution} 58:1403-13.
	\item Brunet E, Rouzine IM, Wilke CO (2008) The stochastic edge in adaptive evolution. \emph{Genetics} 179:603-20.
	\item Cancer T, Atlas G (2012) Comprehensive molecular characterization of human colon and rectal cancer. \emph{Nature} 487:330-7.
	\item Keightley PD, Kryukov G V, Sunyaev S, Halligan DL, Gaffney DJ (2005) Evolutionary constraints in conserved nongenic sequences of mammals. \emph{Genome Res} 15:1373-8.
	\item Kryukov G V, Schmidt S, Sunyaev S (2005) Small fitness effect of mutations in highly conserved non-coding regions. \emph{Hum Mol Genet} 14:2221-9.
	\item Haag-Liautard C et al. (2007) Direct estimation of per nucleotide and genomic deleterious mutation rates in Drosophila. \emph{Nature} 445:82-5.
	\item Tong a H et al. (2001) Systematic genetic analysis with ordered arrays of yeast deletion mutants. \emph{Science} 294:2364-8.
	\item Camps M, Herman A, Loh E, Loeb L a (2007) Genetic constraints on protein evolution. \emph{Crit Rev Biochem Mol Biol} 42:313-26.
	\item Geiler-Samerotte KA et al. (2011) Misfolded proteins impose a dosage-dependent fitness cost and trigger a cytosolic unfolded protein response in yeast. \emph{Proc Natl Acad Sci U S A} 108:680-5.
	\item Boyko AR et al. (2008) Assessing the evolutionary impact of amino acid mutations in the human genome. \emph{PLoS Genet} 4:e1000083.
	\item Bozic I et al. (2010) Accumulation of driver and passenger mutations during tumor progression. \emph{Proc Natl Acad Sci U S A} 107:18545-50.
	\item Howlader N, Noone AM, Krapcho M, Garshell J, Neyman N, Altekruse SF, Kosary CL, Yu M, Ruhl J, Tatalovich Z, Cho H, Mariotto A, Lewis DR, Chen HS, Feuer EJ CK (eds) (2013) SEER Cancer Stastistics Review (CSR) 1975-2010. \emph{EER Cancer Stat Rev 1975-2010, Natl Cancer Institute.} Available at: \url{http://seer.cancer.gov/csr/1975_2009_pops09/results_single/sect_01_table.10_2pgs.pdf}.
	\item Stephens PJ et al. (2012) The landscape of cancer genes and mutational processes in breast cancer. \emph{Nature} 486:400-4.
	\item Berger MF et al. (2012) Melanoma genome sequencing reveals frequent PREX2 mutations. \emph{Nature}:1-18.
	\item Ding L et al. (2008) Somatic mutations affect key pathways in lung adenocarcinoma. \emph{Nature} 455:1069-75.
	\item Bota S et al. (2001) Follow-up of bronchial precancerous lesions and carcinoma in situ using fluorescence endoscopy. \emph{Am J Respir Crit Care Med} 164:1688-93.
	\item Tong WWY et al. (2013) Progression to and spontaneous regression of high-grade anal squamous intraepithelial lesions in HIV-infected and uninfected men. \emph{AIDS} 27:2233-43.
	\item Page DL, Dupont WD, Rogers LW, Rados MS (1985) Atypical hyperplastic lesions of the female breast. A long-term follow-up study. \emph{Cancer} 55:2698-708.
	\item Haigh J (1978) The accumulation of deleterious genes in a population---Muller's ratchet. \emph{Theor Popul Biol} 267:251-267.
	\item Desai MM, Nicolaisen LE, Walczak AM, Plotkin JB (2012) The structure of allelic diversity in the presence of purifying selection. \emph{Theor Popul Biol} 81:144-57.
	\item Shaw AT, Engelman J a (2013) ALK in lung cancer: past, present, and future. \emph{J Clin Oncol} 31:1105-11.
	\item Ciriello G et al. (2013) The molecular diversity of Luminal A breast tumors. \emph{Breast Cancer Res Treat} 141:409-20.
	\item Chou T (2006) Theoretical basis, experimental design, and computerized simulation of synergism and antagonism in drug combination studies. \emph{Pharmacol Rev} 58:621-81.
	\item Jordan DM, Ramensky VE, Sunyaev SR (2010) Human allelic variation: perspective from protein function, structure, and evolution. \emph{Curr Opin Struct Biol} 20:342-50.
	\item Vesely MD, Schreiber RD (2013) Cancer immunoediting: antigens, mechanisms, and implications to cancer immunotherapy. \emph{Ann N Y Acad Sci} 1284:1-5.
	\item Wang ZC et al. (2012) Profiles of genomic instability in high-grade serous ovarian cancer predict treatment outcome. \emph{Clin Cancer Res} 18:5806-15.
	\item Rossi-Fanelli A, Cavaliere R, MondovÏ B, Moricca G (1977) \emph{Selective Heat Sensitivity of Cancer Cells} (Springer Berlin Heidelberg, Berlin Heidelberg).
	\item Goyal S et al. (2012) Dynamic mutation-selection balance as an evolutionary attractor. \emph{Genetics} 191:1309-19.
	\item Bell G, Gonzalez A (2009) Evolutionary rescue can prevent extinction following environmental change. \emph{Ecol Lett} 12:942-8.
	\item Mermel CH et al. (2011) GISTIC2.0 facilitates sensitive and confident localization of the targets of focal somatic copy-number alteration in human cancers. \emph{Genome Biol} 12:R41. 
\end{enumerate}
\clearpage

\renewcommand{\thetable}{S\arabic{table}}   
\renewcommand{\thefigure}{S\arabic{figure}}

\setcounter{figure}{0}
\setcounter{equation}{0}
\makeatletter 
\def\tagform@#1{\maketag@@@{\textbf{[S\ignorespaces#1\unskip\@@italiccorr]}}}
\makeatother

\section{Supplementary Information to A tug-of-war between driver and passenger mutations in cancer and other adaptive process}

\subsection{Mathematical description and analysis of our model of advantageous drivers and deleterious passengers.} 
In this section, we present an exact mathematical formulation of our model, describe the broad ranges of parameters that we chose to explore, and offer an analytical description of our model. In our analytical description, we estimate (i) the effects of stochasticity on population dynamics, (ii) the rate of accumulation of deleterious passengers, and (iii) the interference of driver accumulation by deleterious passengers.

\subsubsection{Detailed formulation of our model} As mentioned in the main text, we model cancer via a first-order Gillespie Algorithm. Each cell within the cancer is represented by a separate ``chemical species" or reactant in the Gillespie algorithm. Cells are defined by their state: $\{n_d, n_p\}$ 
\footnote{Depending upon the number of cells and genomes in the population, it may be more efficient to model cancer as a set genotypes that can gain and lose cells (rather than a set of cells that can gain mutations---as we have done). This alternate design is best when the number of genomes in the population is significantly less than the number of cells. However, the efficiency of a Gillespie algorithm depends very weakly on the number of chemical species $n$ (Specifically, it affects computational speed by $O(log\ n)$ in the Next Reaction Method \cite{Gibson2000}). More importantly however, the other steps in the simulation: creating mutations, and calculating birth/death rates is faster for individual cells than for genomes. Thus, using cells as the basic element of simulations, rather than genomes, is faster even under circumstances where the number of cells outnumber the number of genomes by a few orders of magnitude. This design choice also afforded more plasticity in model design and allowed us to create and investigate coalescences trees, which were informative. In fact, the genetics community at large tends to simulate genomes (rather than individuals) for what appears to be ill-conceived speed considerations, and may want to consider adopting our style. }. 
$n_d$ denotes the number of drivers in the cell, while $n_p$ denotes the number of passengers. Cells can then divide, with and without mutations, and die according to the following half-reactions:
\begin{equation*}
  \begin{array}{lcc}
\{n_d, n_p\} 	& \xrightarrow{B(n_d, n_p)}	& \{n_d + \delta d, n_p + \delta p\} + \{n_d + \delta d, n_p + \delta p\}			\\
\{n_d, n_p\} 	& \xrightarrow{D(N)}					& \emptyset				\\
  \end{array}
\end{equation*}
The functions $B(n_d, n_p)$ and $D(N)$ represent the birth and death rates of cells, while $N$ represents the total number of cells in the precancerous population. The birth rate assumes multiplicative fitness effects of mutations and no epistasis between mutations:
\begin{equation}\label{eq:birth}
B(n_d, n_p) = \frac{(1+s_d)^{n_d}}{(1+s_p)^{n_p}} \approx (1+s_d)^{n_d}(1-s_p)^{n_p}
\end{equation}
We also define a generation in terms of the mean division time:
\begin{equation*}
\mbox{1 generation} = \frac{1}{1/N \sum_{i=0}^N B(n_{d_i}, n_{p_i})} 
\end{equation*}

The death rate is defined such that, in the absence of mutations, the expectation value of the population size will obey a Gompertz curve at large sizes and a logistic curve at small sizes:
\begin{equation*}
D(N) = \mbox{Log}[1+\frac{(e-1)N}{N^0}] 
\end{equation*}
We used a simpler form of this death function for populations grown to less than $10^6$ cells:
\begin{equation}\label{eq:death}
D(N) = \frac{N}{N^0}
\end{equation}
This second functional form did not significantly alter dynamics at small sizes \cite{McFarland2013}, has been used previously \cite{Johnston2007}, is easier to calculate, and seemed equally justified to us for small sizes because very little is known about the true carrying capacity of a tumor micro-environment in its early stages.  
Lastly, the number of new drivers $\delta d$ and passengers $\delta p$ acquired during cell division are Poisson-distributed random variables with mean $U_d$ and $U_p$, respectively. For example, $P(\delta p = k | U_p) = \frac{U_p^k e^{-U_p}}{k!}$. 

Many of the particular design choices and properties of our model were altered and then investigated in a previous study \cite{McFarland2013}. Specifically, we considered (1) the effects of mutations with additive effects [i.e. $B(n_d, n_p) = 1 + n_d s_d - n_p s_p$], (2) the effects of mutations that alter the death rate [i.e. $D = D(N, n_d, n_p)$], (3) the effects of driver and passenger mutations selected from various distributions (exponential, normal, and gamma) of fitness effect sizes, and (4) variations on the relation between population size and death rate. For the parameters that we believe are most relevant to cancer (\textbf{Table \ref{table:parameters}}), these permutations did not qualitatively alter our simulations. However, in the analytical analysis presented below we discuss the boundaries where assumptions of our model break-down; this was, in part, why we analyzed the model in such detail.

It should be noted that many of the considerations discussed above are germaine to all models of tumor progression, not simply the \emph{in silico} model presented here. Consider that recent data on growth rates of human tumors differs from data obtain from mouse models: human tumors grow according to an exponential curve \cite{Bota2001}, while mouse tumors grow according to a Gompertz curve \cite{Domingues2012}. Careful mathematical consideration of the differences between a model of progression where growth is exponential, and one where growth is Gompertzian, should allow us to understand when it is necessary to refine the design of simulations and experiments and/or temper our conclusions. 
	
Before describing the entire dynamics of our model, it is useful to consider the difference between our simulations initiated at their stationary size ($N^0$ cells) and simulations initiated at 1 cell. In the absence of mutations, an initial population of one cell will grow logistically until it reaches the stationary size. Hence, it takes approximately $\mbox{Log}_2[N^0] \sim \mbox{Log}_2[10^3] \sim 10$ generations for the initial cell to approach stationary size. This is far shorter than the average time required for cancer progression ($\sim 10,000$ generations) and the time required for a new driver to accumulate ($\sim 1 /(U_dN^0s_d) \sim 1,000$ generations). Thus, our choice of initiating a tumor at one cell versus $N^0$ does not significantly alter the conclusions of our model. 

This comparison of timescales also suggests that cancers are almost always near their stationary size:
\begin{equation*}
\overline{B(n_d, n_p)} \approx D(N) 										\\
\end{equation*}
We previously tested this conclusion in simulations and found that it is a excellent approximation of tumor size \cite{McFarland2013}. If we assume $\overline{B(n_d, n_p)} \approx B(\overline{n_d}, \overline{n_p})$, then a relationship between the number of drivers and passengers in a tumor and its size is obtainable:
\begin{equation*}
  \begin{array}{c}
B(\overline{n_d}, \overline{n_p}) \approx D(N) \\
\frac{(1+s_d)^{\overline{n_d}}}{(1+s_p)^{\overline{n_p}}} \approx \mbox{Log}[1 + \frac{N}{(e-1)N^0}] 	\\
\overline{n_d} \mbox{Log}(1 + s_d) - \overline{n_p} \mbox{Log}(1 + s_p) \approx
\mbox{Log}[\mbox{Log}[\frac{N}{N^0}]] \\
\overline{n_d}s_d - \overline{n_p}s_p \approx 
\mbox{Log}[\mbox{Log}[\frac{N}{N^0}]] :
s_d, s_p \ll 1	\\ 
 \end{array}
\end{equation*}
This final equation suggests that there exist a linear relationship between drivers and passengers among tumors with similar $s_d$ and $s_p$, which we assume is the case for tumors of the same tissue of origin. The relationship should be relatively robust to tumor size, but sensitive to the fitness effects of drivers and passengers. Moreover, changes in the functional form of $D(N)$ will alter the y-intercept of this linear relationship, but not the slope of the relationship. Hence, we can draw conclusions about the relative strength of drivers versus passengers ($s_d/s_p$) without knowing the exact constraints on population size. We tested and verified this prediction of a linear relationship between drivers and passengers in the main text. 

Our computational model has 5 independent parameters: a mutation rate $\mu$, a mutation's relative likelihood of being a driver versus a passenger $T_d/T_p$, the fitness benefit of a driver $s_d$, and disadvantage of a passenger $s_p$, and an initial stationary size $N_0$. These parameters vary considerably between tumor types (and the mutation rate even varies within tumor types \cite{Lawrence2013}), so we explored a wide range of values centered around literature best-estimates (\textbf{Table \ref{table:parameters}}). More importantly, our analytical analysis reveals that we can describe our system with two dimensionless parameters, which we then estimated from age-incidence and genomics data (\textbf{Fig. 2}).

\subsubsection{Overview of our analytical model for dynamics} 
In the main text, we demonstrate that dynamics are described by two countervailing forces: an upward velocity $v_d$ resulting from accumulating beneficial drivers, and a downward velocity $v_p$ resulting from accumulating deleterious passengers. The upward velocity $v_d$ was further subdivided into a product of the rate at which new drivers fixate in the population $f$ times their effect on population size once fixated $\Delta N$ (\textbf{Fig. 1B})
\footnote{While we assume that drivers arise at random time intervals, this assumption is not always true. Because unfixed passengers can interfere with the fixation of drivers, a driver is more likely to fixate immediately following a previous driver fixation event \cite{Johnson2002}. Ignoring this caveat does not significantly alter dynamics in the parameter space explored here.}. 
The velocities $v_d$ and $v_p$ are balanced at a critical population size $N^*$, at which the population is approximately equally likely to go extinct or progress to cancer. 

While we were able to describe the average behavior of our population in the main text, our system, like cancer, is inherently stochastic. Its complete dynamics is best described by a differential equation with stochastic jumps:
\begin{equation}\label{eq:diff}
  \begin{array}{c}
dN = v_p dt + \Delta N dn_d 	\\
n_d \xrightarrow{f} n_d + 1			\\
  \end{array}
\end{equation}
In this equation, the change in population size $dN$ is the product of a deterministic component $-v_p$, along with a stochastic component describing the random arrival of new drivers $(\Delta N dn_d)$. Below, we use this equation to estimate the probability of cancer for any population size $\Pcancer(x)$ and the mean waiting time to cancer $\overline{\Tcancer}(x)$. Lastly, we noticed that simulations differed from the formalism we presented in the main text when we varied the mutation rate $\mu$ and explored a broader range of passenger deleteriousness $s_p$ (\textbf{Fig. 2, \ref{fig:phase_space}}). These discrepancies could be resolved by considering two phenomena neglected by our first derivation: selection against passengers, and passenger's effect on both the fixation probability and clone fitness of drivers. Fortuitously, accounting for these phenomena did not alter Eq. \ref{eq:diff}, nor the overall framework of our analytical model. Instead, they only affect the rates $v_p$, $f$, and jump size $\Delta N$ in our model. Thus, with the refined formalism, we described dynamics across a very broad range of parameters (\textbf{Fig. 2, \ref{fig:phase_space}}). More importantly, we observe drastic reductions in the probability of adaptation at high mutation rates and encompassing moderately deleterious passengers. These findings suggest novel strategies to cancer therapy. 

Population size is the state variable of our system and, as such, is all that is needed to describe future dynamics (this is evident from Eq. \ref{eq:diff}, and can be observed in \textbf{Fig. \ref{fig:path_independence}}). By converting population size into a dimensionless parameter $x = N/N^*$ (and $x^0 = N^0/N^*$), the probability of cancer collapse onto a simple curve $\Pcancer(x)$ (\textbf{Fig. 1})---further underscoring the importance of the critical population size. Hence, we will use this dimensionless quantity heavily throughout the remainder of our analysis.

\subsubsection{Estimating the probability of cancer}
Using Eq. \ref{eq:diff} we can describe how the probability of extinction changes in an infinitesimal time due to either passenger accumulation or a rare driver jump: 
\begin{equation*}
\Pcancer(x) = f(x) dt \Pcancer[x + \Delta N(x)] 
+ [1 - f(x) dt] \Pcancer[x - v_p(x) dt]
\end{equation*}
Note that $f$, $\Delta N$, and $v_p$ are all functions of $x$. In this equation, we see that is the probability of cancer at $x$ is the probability of a jump times the probability of cancer after the jump ($f(x) dt \Pcancer[x + \Delta N(x)]$) plus the probability of decline times the probability of cancer after the decline ($[1-f(x) dt] \Pcancer[x - v_p(x) dt]$). The probability of cancer after a decline can be expanded via a Taylor series: $\Pcancer[x - v_p(x) dt] \approx \Pcancer(x) - v_p dt P'_{\rm{cancer}}(x)$. This reduces the above equation to:
\begin{equation}\label{eq:pre_maclaurin}
v_p(x)P'_{\rm{cancer}}(x) = f(x) [ \Pcancer(\theta x) - \Pcancer(x) ] 
\end{equation}
Here $\theta = (x + \Delta N)/ x \approx 1 + s_d$ denotes the logarithmic change in population size after a driver jump. Like $\Delta N$, $v_p=\mu_p s_p N$ and $f=\mu_d s_d N$ are also linear in $x$. Thus, by expanding the logarithm of $\theta$ via Maclaurin Series: $\Pcancer[ x + \Delta N(x)] \approx \Pcancer(x) + x \mbox{Log}(\theta) P'_{\rm{cancer}}(x) + x^2 \mbox{Log}^2(\theta)  + P''_{\rm{cancer}}(x) $, we arrive at a now solvable differential equation: 
\begin{equation*}
x^2 \mbox{Log}(\theta) P''_{\rm{cancer}}(x) + [\mbox{Log}(\theta) x + 2 x + 2 N^* ] P'_{\rm{cancer}}(x) = 0
\end{equation*}
With this equation, and boundary conditions that follow from our definition of cancer and extinction:
\begin{equation*}
  \begin{array}{r}
\Pcancer(x=0) = 0 \\
\Pcancer(x=\infty) = 1 \\
  \end{array}
\end{equation*}
, we can solve for the probability of cancer after infinite time:
\begin{equation}\label{eq:Pcancer}
  \begin{array}{c}
\Pcancer = 1 - \gamma(\frac{2}{\mbox{Log}(\theta)}, \frac{2}{\mbox{Log}(\theta)x})
  \end{array}
\end{equation}
Here, $\gamma(s, x) = 1/ \Gamma(s) \int_0^x e^{-t} t^{s-1} dt : \Gamma(s) = \int_0^\infty  x^{s-1} e^{-x}\,{\rm d}x$ is the normalized incomplete gamma function. This solution is parameterized by two dimensionless quantities: $\theta$ and $x$, which represent the jump size in population of driver sweeps and our effective population size respectively.

\subsubsection{Estimating the mean time to progression}					
We can also use Eq. \ref{eq:diff} to solve for the waiting time to cancer. This can be accomplished in two ways: (1) we can simulate random driver jumps and deterministic passenger decline directly, and (2) we can approximate the mean waiting time to cancer using a Taylor expansion similar to the strategy we employed to solve for the probability of cancer. These two approaches agree with each other (thus, illustrating their accuracy), and offer key insights into the evolutionary parameters that affect age-incidence curves (\textbf{Fig. 2, \ref{fig:incidence}}).

Eq. \ref{eq:diff} can be simulated using a ``hybrid'' Gillespie algorithm: a meta-simulation of driver- and passenger-accumulation events that we, originally, observed arising from our atomistic simulations of birth, death, and mutational events. The advantage of this technique is that it allows us to quickly simulate billions of tumors, which would be computationally impossible via the more detailed simulations. Because we are confident that we are accurately estimating the rate of driver and passenger accumulation events (\textbf{Fig. \ref{fig:phase_space}}), this simplification should retain accuracy. To simulate Eq. \ref{eq:diff} directly, we must consider that the instantaneous probability of a driver jump $f[x(t)]$ is a function of a constantly declining population size due to passenger accumulation: $x(t) = x_{n_d}(1+s_p)^{v_p/s_p t} \approx x_{n_d}e^{-v_p t}$. Here, $x_{n_d}$ is the population size after the last driver jump. Thus, the waiting time between drivers $\Delta t = t_{n_d+1}- t_{n_d}$ is:
\begin{equation}
  \begin{array}{c}
\int_0^{\Delta t} f N(t') dt' 			= \zeta \\
f \int_0^{\Delta t} N_{n_d} e^{-v_p t'} dt' = \zeta \\
\Delta t = -\frac{1}{v_p} \mbox{Log}(1- \frac{v_p \zeta}{f N_{n_d}}) \\
  \end{array}
\label{eq:hybrid_gillespie}
\end{equation}
, where $\zeta$ is an exponentially-distributed random number with mean 1. Using our precise calculations of $f$, $v_p$ and $\Delta N$ below, we can now simulate Eq. \ref{eq:diff} directly. 

We can also solve Eq. \ref{eq:diff} for $\Tcancer$, using the exact same approximations as we did to estimate $\Pcancer(x)$. To do this, we begin with a Master Equation for the probability of acquiring a cancer after waiting time $t$ when currently at size $x$: 
\begin{equation*}
\Pcancer(x,t)=f(x) \delta t\ \Pcancer(\theta x,t+\delta t) 
+[1-f(x)\delta t]\Pcancer[x - v_p(x) \delta t, t + \delta t]
\end{equation*}
The mean waiting time to cancer is then:
\begin{equation*}
\overline{\Tcancer(x)}=\int _0^{\infty} t \Pcancer(x,t) dt
\end{equation*}
By substituting the Master Equation into this definition, and by utilizing the first-order Taylor series expansion:
\begin{equation*}
\Pcancer(\theta x,t - \delta t) \approx \Pcancer(x, t) + 
						   \frac{\partial \Pcancer(x, t)}{\partial x} (\theta - 1) + \frac{\partial \Pcancer(x, t)}{\partial t} \delta t 
\end{equation*}
, we find:
\begin{equation*}
\overline{\Tcancer}(x) 		= \int_0^\infty \Pcancer(x, t) t dt \\
						+ \delta t \int_0^\infty \frac{\partial \Pcancer(x, t)}{\partial t} t dt \\
						+ [ f \delta t (\theta-1) + \big(1- f \big) \big(v_p \delta t \big) ] \int_0^\infty \frac{\partial \Pcancer(x, t)}{\partial x} t dt
\end{equation*}
The first integral in this solution is simply the definition of our mean waiting time ($\overline{\Tcancer}(x)$). The second integral can be integrated by parts by noting that $\lim_{t \to \infty} t \Pcancer(x, t) = 0$ (otherwise, $\overline{\Tcancer}(x)$ would be undefined). Lastly, the third integral reduces to $\overline{t'_{\rm{cancer}}}(x)$. Thus, we eventually find: 
\begin{equation*}
f(x)[\rho_c (\theta x)- \rho_c(x)] - v_p(x) \rho_c'(x) + \Pcancer(x) = 0 
\end{equation*}
Here, $\rho_c(x) = \Pcancer(x) \overline{\Tcancer}(x)$. This equation has a nearly identical form to Eq. \ref{eq:pre_maclaurin}. So we used a similar Second-Order Maclaurin series expansion of $\theta$ to approximate its solution:
\begin{equation}
\label{eq:progression_time}
\overline{\Tcancer}(x) = \frac{2}{f \mbox{Log}^2(\theta)}\Big[
\int_x^\infty \frac{dy}{y^3} \frac{\Pcancer(y)[1 - \Pcancer(y)]}{P'_{\rm{cancer}}(y)} \\
+ \frac{1 - \Pcancer(x)}{\Pcancer(x)} \int_0^x \frac{dy}{y^3} \frac{P^2_{\rm{cancer}}(y)}{P'_{\rm{cancer}}(y)} \Big]
\end{equation}
This equation can be solved using Simpson's Method and is in good agreement with the hybrid simulations described in the preceding paragraph (\textbf{Fig. \ref{fig:incidence}}). This calculation of the waiting time to cancer is most illustrative when $x \ll 1$---the regime that we expect to contain most tumors. In this regime, the mean time increases as $-\mbox{Log}(x)/v_p$, which implies two interesting properties of $\overline{\Tcancer}$. First, $x$ has a very weak, sub-linear, effect on the waiting time and does not significantly alter the shape of incidence curves (\textbf{Fig. \ref{fig:incidence}}). Second, the waiting time to cancer is dictated by $v_p$ (the accumulation rate of passengers), thus offering yet another reason to understand the rate of deleterious passenger accumulation.

\subsubsection{Accumulation of deleterious passengers} 								
Passenger mutations accumulate and drag populations down by a rate $v_p$. This quantity is a product of passengers arrival rate $\mu_p N$, their fixation probability $\pi_p$, and their effect on population size once fixated $N s_p$ (i.e. $v_p \approx \mu_p s_p N$). In the main text, we assume that the fixation probability is approximately neutral ($\pi_p \approx 1/N$); however, when selection is stronger that genetic drift, the fixation probability becomes less than the neutral rate. A number of studies have focused entirely on estimating this fixation probability \cite{Neher2012, Goyal2012, Brunet2008, Gordo2000}. In general, estimates of this fixation probability begin by considering the distribution of deleterious alleles in a population of infinite size in mutation-selection balance---where allele frequencies are not changing. At equilibrium, such a population exhibits a Poisson distribution in the number of segregating passengers within cells $N_{n_p}$, defined by a characteristic parameter $\lambda_p = \mu_p/s_p$ (\textbf{Fig. 3C}): 
\begin{equation}\label{eq:mutation-selection_balance}
N_{n_p} = N \frac{e^{-\lambda_p} \lambda_p^{n_p}}{n_p!}
\end{equation}
If we then consider a population of finite size, we find that the allele frequencies fluctuate due to genetic drift. If fluctuations in the fittest class ($N_{n_p=0} = N e^{-\lambda_p}$) are large enough to cause this fittest class to go extinct, then it is irrevocably lost from the population. This irrevocable loss is considered a `click' of Muller's Ratchet. The new fittest class---individuals harboring one segregating passenger prior to the `click'---then relaxes to a new equilibrium that fluctuates, and the process repeats. Estimating the time required for a new fittest class to relax to equilibrium size immediately following a `click' is non-trivial and dependent upon the parameters of the system: $N$, $s_p$, and $\mu_p$, which can vary by orders of magnitude depending upon the evolutionary system in question; hence there are many estimates of the exact rate of Muller's Ratchet. We present and utilize 3 estimates of the rate of Muller's Ratchet: (1) a solution that ignores the time to equilibration and works decently for most values of $s_p$, $\mu_p$, and $N$ considered here (\textbf{Fig. 2, magenta lines}); (2) a traveling-wave solution that allows the distribution of segregating passengers to be far from equilibrium, but presumes that that the size of neighboring fitness classes are uncorrelated, accurate for large values of $\lambda_p$ \cite{Brunet2008}; and (3) a path-integral solution that considers correlations between neighboring fitness classes, but requires that the population be in quasi-equilibrium (i.e. near mutation-selection balance), accurate for small values of $\lambda_p$ \cite{Gordo2000}. These later two estimates were combined with an estimate of the number of hitchhiking passengers and their effects on driver fixation events to form a very precise description of our model's dynamics (\textbf{Fig. 3, \ref{fig:phase_space}; black lines}). We felt that offering these two refined models to readers was useful because they trade-off applicability, simplicity, and accuracy. 

If we simply ignore the equilibration tome of a population into mutation-selection balance, then we can estimate the rate of Muller's Ratchet with a closed form solution that is applicable to all values of $s_p$, $\mu_p$, $N$ investigated here. The other two solutions each apply only to their respective halves of the parameter space and are more complex, but also more accurate. To our knowledge, we are the first authors to present this simplified solution. We obtain the solution by assuming that the probability of a `click' is approximately the probability of a new passenger fixating \emph{within} the fittest class: $N_{n_p = 0} = N e^{-\lambda_p}$. In other words, to a first-approximation, deleterious passengers simply reduce the effective population size of our system. The probability of a lone deleterious allele fixating within this fittest class is describe by a Moran Process \cite{nowak2006evolutionary}. Hence,
\begin{equation}\label{eq:kirill_ratchet}
\pi_p^{(1)} = \frac{s_p}{(1 + s_p)^{N_{p_0}} - 1}
\end{equation}
This refined fixation probability $\pi_p^{(1)}$ is then used to correct the downward velocity due to passengers, using the same formula for $v_p$ derived in the main text: 
\begin{equation}\label{eq:generalized_vp}
v_p^{(i)} = \mu_p s_p N_{p_0} \pi_p^{(i)}
\end{equation}
This equation links $v_p$ to the passenger fixation probabilities calculated above, and the other two fixation probabilities calculated below. 

The solution for Muller's Ratchet as a traveling wave, which we apply when $\lambda_p < 1$, was obtained from \cite{Brunet2008}:
\begin{equation}
\label{eq:wave}
\frac{\mbox{Log}(\frac{N s_p}{\sqrt{\lambda_p}})}{\lambda_p} \approx 
1 - \frac{\pi_p^{(2)}}{2}[\mbox{Log}^2(\frac{e}{\pi_p^{(2)}}) + 1] \\
- \frac{1}{\lambda_p} \mbox{Log}[\frac{(\pi_p^{(2)})^{3/2}}{\sqrt{1-pi_p^{(2)}}}\frac{\mbox{Log}(\frac{e}{\lambda_p})}{1 - \pi_p^{(2)} \mbox{Log}(\frac{e}{\lambda_p}) + \frac{5}{6 \lambda_p}}]
\end{equation}
Because this equation is transcendental, we solved for $\pi_p^{(2)}$ using Brent's Method. 

When $\lambda_p \ge 1$, a quasi-stationary analysis of the mutation classes becomes appropriate. This analysis was first done in \cite{Gordo2000}, resulting in a solution of the form:
\begin{equation}\label{eq:quasi-stationary}
T_{click} = \frac{e-1}{s_p}e^{\frac{s_p N_{p_0}}{2(e-1)}}
\end{equation}
The fixation probability is then simply the inverse of the `click' time: $\pi_p^{(3)}=1/T_{click}$.

Lastly, there is a discontinuity between the above two solutions at their intersection: $\lambda_p = 1$. We resolved this by interpolating between the two solutions, as follows:
\begin{equation*}
\pi_p^{(\rm{combined})} = \lambda_p \pi_p^{(2)} + (1 - \lambda_p) \pi^{(3)}
\end{equation*}

\subsubsection{Effects of deleterious passengers on fixation probability and clone fitness of drivers}
The occurrence and fixation of driver mutations are rare events, separated by nearly random time intervals, with a frequency of occurrence $f = \mu_d N \pi_d$. Here, $\pi_d$ is the fixation probability of a new mutant driver once it arises in the population. In the first-order model presented in the main text, we estimate that $\pi_d = s_d/(1+s_d) \approx s_d$. However, this result assumes that there are no other non-neutral alleles in the population. In reality, there are many segregating passengers in the population, and potentially other segregating drivers. 

The presence of other drivers in the population, which interfere with the fixation of our clone of interest, is a phenomena commonly described as \emph{Clonal Interference} \cite{Good2012a}. Clonal Interference becomes significant in the population once the time required for a driver to fixate [$\sim \mbox{Log}(N)/s_d$ generations] approaches the fixation rate ($f \approx \mu_d N s_d$). Nascent precancerous population are in a space of evolutionary parameters where Clonal Interference is particularly negligible: population size is small ($N \sim 10^3$), and drivers are rare ($\mu_d \sim 10^{-5}$), but strong ($s_d \sim 10^{-1}$). Thus, we do not consider its effects here. However, for a larger tumor population, clonal interference may become very significant. This is especially true in a poorly-mixed population, where beneficial alleles take longer to sweep through the population \cite{Korolev2012a}. 

\newcommand{\pI}[0] {\delta p_{\rm{I}} }
\newcommand{\pS}[0] {\delta p_{\rm{S}} }

Segregating passenger mutations can also interfere with a driver sweep by `hitchhiking' on the expanding clone \cite{Johnson2002, Bachtrog2004}. We consider two types of hitchhikers: (1) those that reside in the \underline{I}nitial clone before the new driver arises (denoted $\pI$), and (2) those that arise and fixate in the new driver clone as it \underline{S}weeps through the population (denoted $\pS$).  It is necessary to distinguish hitchhikers this way because only the initial hitchhikers ($\delta p_I$) significantly alter the fixation probability $f$, while both types alter the effect size $\Delta N$.  The hitchhikers that accumulate during the sweep will generally arise after the clone is of appreciable size; however, once the driver clone is of appreciable size, it is exceedingly likely that it will fixate so long as it remains the fittest clone in the population. 

Here, we consider only the average number of hitchhikers in a driver sweep ($\overline{\pI}$ and $\overline{\pS}$), rather than their entire distribution of quantities; estimates of the average number of hitchhikers appear to explain dynamics reasonably well (first shown in \cite{Johnson2002} and also evident from our analysis' good agreement with simulations \textbf{Fig. \ref{fig:phase_space}}). Thus the probability that a new clone fixates in the absence of Clonal Interference is (\textbf{Fig. 3C}):
\begin{equation}\label{eq:driver_fixation}
\pi_d(\overline{\pI}) = \frac{s_d'(\overline{\pI})}{1+s_d'(\overline{\pI})} : s_d'(\overline{\pI}) = s_d - \overline{\pI} s_p
\end{equation}
, and the jump size $\Delta N$ becomes:
\begin{equation}\label{eq:jump_size}
\Delta N' = N [s_d - (\overline{\pI} + \overline{\pS}) s_p]
\end{equation}
We can conclude our analysis of hitchhikers once we obtain $\overline{\pI}$ and $\overline{\pS}$. These quantities were first derived in \cite{Johnson2002}. We use their results (summarized below), along with a minor necessary adjustment for populations when $\lambda_p$ is large, to complete our analytical model of cancer progression. 

For a new driver clone to take over the population and fixate, it has been shown that its fitness must be greater than the fittest class in the population \cite{Johnson2002}. This imposes a maximum on the number of initial hitchhikers $\pI^{\rm{max}}$ that a successful driver clone can have:
\begin{equation*}
  \begin{array}{rcl}
s_d 				&	>	&	\pI s_p 				\\
\pI^{\rm{max}} 	&	=	& \lfloor s_d/s_p \rfloor 	\\ 
  \end{array}
\end{equation*}
A clone that does not satisfy this constraint may proliferate for a while in the population, but it will nevertheless be eventually out-competed by fitter clones. When the mean number of hitchhiking passengers ($\lambda_p$) approaches this maximum, hitchhikers dramatically reduce both $f$ and $\Delta N$, thus increasing $N^*$ to untenable sizes. This occurs when:
\begin{equation}\label{eq:critical_mu}
  \begin{array}{rcl}
\lambda_p 	&	=		& \pI^{\rm{max}}		\\
\mu_p/s_p 	&	= 		& \lfloor s_d/s_p \rfloor	\\
\mu_p 		&\approx  & s_d 						\\
  \end{array}
\end{equation} 
Hence, our analysis suggests a limit on the maximum mutation rate that an adapting population can tolerate: $\mu_p^* \approx s_d$. In simulations, we observe extinction slightly above this threshold (\textbf{Fig. 3A, S2}). This mechanism of collapse, where populations go extinct by failing to acquire new advantageous mutations or adaptations, differs from the traditional model of mutational meltdown. In the traditional model, advantageous mutations are generally ignored and meltdown occurs only because deleterious mutations accumulate too quickly. In our model, however, traditional mutational meltdown is difficult because populations also acquire advantageous mutations faster as the mutation rate increases. Moreover, traditional meltdown occurs only when the population size is small, making it impossible to occur in a large population like cancer. Our discovery of a new mechanism of meltdown that is independent of population size suggests that mutational meltdown may be induced via cancer therapeutics.

The number of initial segregating passengers in a clone when a driver arises ($\delta p_I$) can be obtained by considering, once again, the population at mutation selection balance, i.e. Eq. \ref{eq:mutation-selection_balance}. The average number of initial hitchhiking passengers is simply the average of the likelihood of a driver arising in each mutational class, conditional on the driver successfully sweeping through the population:
\begin{equation}\label{eq:initial_hitchhikers}
  \begin{array}{rcl}
P(\pI) 			&=&	\frac{1}{\mathscr{N}} N_{n_p=\pI} \pi_d(\pI)  					\\
\overline{\pI}	&=& 	\frac{1}{\mathscr{N}} \sum_{\pI=0}^{\pI^{\rm{max}}} P(\pI) \pi_d(\pI)	\\ 
				&=&	\frac{1}{\mathscr{N}} \sum_{\pI=0}^{\pI^{\rm{max}}} \frac{e^{-\lambda_p} \lambda_p^\pI}{\pI!}   \frac{s_d'(\pI)}{1 + s_d'(\pI)}		\\
  \end{array}
\end{equation}
Here, $\mathscr{N} = \sum_{\pI=0}^{\pI^{\rm{max}}} \pi_d'(\pI)$ is a normalization constant. 

The above solution fails when $\lambda_p$ is large. In this circumstance, the population is far from mutation-selection balance. Rectifying the solution in this case is difficult to do precisely, however we find that a simple correction to Eq. \ref{eq:initial_hitchhikers} can crudely ameliorate the estimate. Because the assumption of mutation-selection balance fails only once the expected number of passengers in the fittest class becomes very small ($N_{n_p=0}=Ne^{-\lambda_p} \sim 1$), we propose that the actual fittest surviving class in the population is the first class of passengers with an expected population size that is greater than the size of fluctuations in the population. Because the variance in a birth and death process is the sum of the rates ($2N$ in our model), the Fittest Surviving Class $k_{\rm{FSC}}$ is: 
\begin{equation*}
  \begin{array}{rcl}
k_{\rm{FSC}} &=& min_{n_p} [ N_{n_p} > \sqrt{2N} ]									\\
k_{\rm{FSC}} &=& min_{n_p} [ e^{-\lambda_p} \lambda_p^{n_p} /n_p! > \sqrt{\frac{2}{N}} ]	\\
  \end{array}
\end{equation*}
The corrected distribution of $\pI$ then becomes:
\begin{equation*}
\overline{\pI} = \frac{1}{\mathscr{N}} \sum_{\pI = 0}^{\pI^{\rm{max}}}P(k = \pI + k_{\rm{FSC}} | \lambda_p )  \pi_d(\pI)
\end{equation*}
 This simple correct yields a final solution for $\Pcancer$ that agrees with simulations well (\textbf{Fig. \ref{fig:phase_space}}).
 
 Lastly, the number of passengers that accumulate during the selective sweep ($\pS$) can be calculated using a recursive relationship.  that begins with the probability of accumulating the maximum possible passengers during the sweep $\pI^{\rm{max}}$ \cite{Johnson2002}:
\begin{equation*}
  \begin{array}{rcl}
P(\pS = \pI^{\rm{max}}) 	&=& \frac{1}{\mathscr{N}_2} \pI^{max}		 								\\
P(\pS = k) 					&=& \frac{1}{\mathscr{N}_2} \frac{k + s_p P(\pS = k + 1)}{1 + s_p}		\\
\overline{\pS}				&=& \frac{1}{\mathscr{N}_2} \sum_{\pS=0}^{\pI^{\rm{max}}} P(\pS)	\\
  \end{array}
\end{equation*}
Where $\mathscr{N}_2 = \sum_{\delta p_S=0}^{\delta p^{max}} P(\delta p_S)$ is a second normalization constant.

\subsection{A traditional model of cancer progression with drivers and neutral passengers.} 
In the traditional model of cancer progression used to estimate age-incidence curves, it is assumed that a cancerous population transitions through $k$ intermediate states before malignancy:
\begin{equation*}
C_0 \xrightarrow{r_1} C_1 \xrightarrow{r_2} ... \xrightarrow{r_k} C_k
\end{equation*}
Simply put, these intermediate states and transitions correspond to the many phenotypic changes that occur within a tumor as it progresses \cite{Hanahan2011}. The instantaneous probabilities of each transition from one state to the next $r_i$ can vary in the general case. Nevertheless, it has been shown that this predicts similar age-incidence rates to a model where transition rates are all the same \cite{ARMITAGE1954}. Thus, for parsimony we only consider the case where all transition rates are the same constant $r$. Moreover, if the transition rates are drastically different from one another, then dynamics will largely be determined by the slowest rate and the faster rates can be neglected.

From a mathematical perspective, this model is agnostic towards the underlying event that transitions a precancerous population from one state to the next. However from a genetic perspective, each transition corresponds to the acquisition of a new driver alteration in the population. If rate-limiting events are non-heritable, then the inferences we draw from age-incidence curves about this traditional model may break down; however, we would like to reiterate that a large variety of events can be drivers in the sense that they are inheritable across cell division: SNMs, SCNAs, alterations in DNA and histone moieties, stable changes in cell signaling cascades, etc. Therefore, we believe it is reasonable to assume that each rate-limiting step is the acquisition of a new driver, as has been presumed for many years \cite{Meza2008}.

We now consider the properties of this model when neutral passengers also accumulate. These neutral passengers do not alter progression. The precancerous population is now defined by the state $C_{n_d, n_p}$. We consider the case where drivers accumulate at a fixed rate $r_d$ and passengers accumulate at different fixed rate $r_p$:

\begin{equation*}
  \begin{array}{ccccc}
C_{0, 0} 			& \xrightarrow{r_d} 	& C_{1, 0} 		& \xrightarrow{r_d} 	& ... \\
 \downarrow{r_p} 	& 				& \downarrow{r_p}	& 				&	\\
C_{0, 1}			& \xrightarrow{r_d} 	& C_{1, 1} 		& \xrightarrow{r_d} 	& ...	\\
 \downarrow{r_p} 	& 				& \downarrow{r_p}	& 				&	\\
\vdots			& 				& \vdots 			& \ddots 			& 	\\
				&				&				&				& C_{n_d, n_p} \\
  \end{array}
\end{equation*} 
As before, cancer arises once enough drivers accumulate ($C_{n_d = k, n_p}$). 

\newcommand{\tcancer}[0] {\tau_{\rm{cancer}}}

To interpret age-incidence data, as well as genomics data, we are interested in both the waiting time until cancer ($\Tcancer$) and the total number of mutations ($n_p + k$). This model can be simplify by noting that there is a freedom in the units for which we measure time. In our simulations, time was measured in generations and then converted to years. Here, we chose to measure time in units of the driver transition probability $r_d$ and will then convert this to years afterwards. Hence, $r_d = 1$ without loss of generality. Consider the quantity $\tcancer=\Tcancer r_d$, as a dimensionless measure of the waiting time to cancer. Because driver and passenger accumulation events are independent processes in this model, the joint probability of observing a cancer at time $\tcancer$ with $n_p$ passenger mutations, $P(\tcancer, n_p| n_d=k, r_p)$, is:
\begin{equation}\label{eq:joint_tau}
P(\tcancer, n_p| n_d=k, r_p) = P(\tcancer| n_d=k)\cdot P(n_p| \tcancer, r_p )
\end{equation}
This joint probability distribution provides a framework for identifying our quantities of interest. 

\begin{figure}
\begin{center}
\centerline{\includegraphics[width=0.75\textwidth]{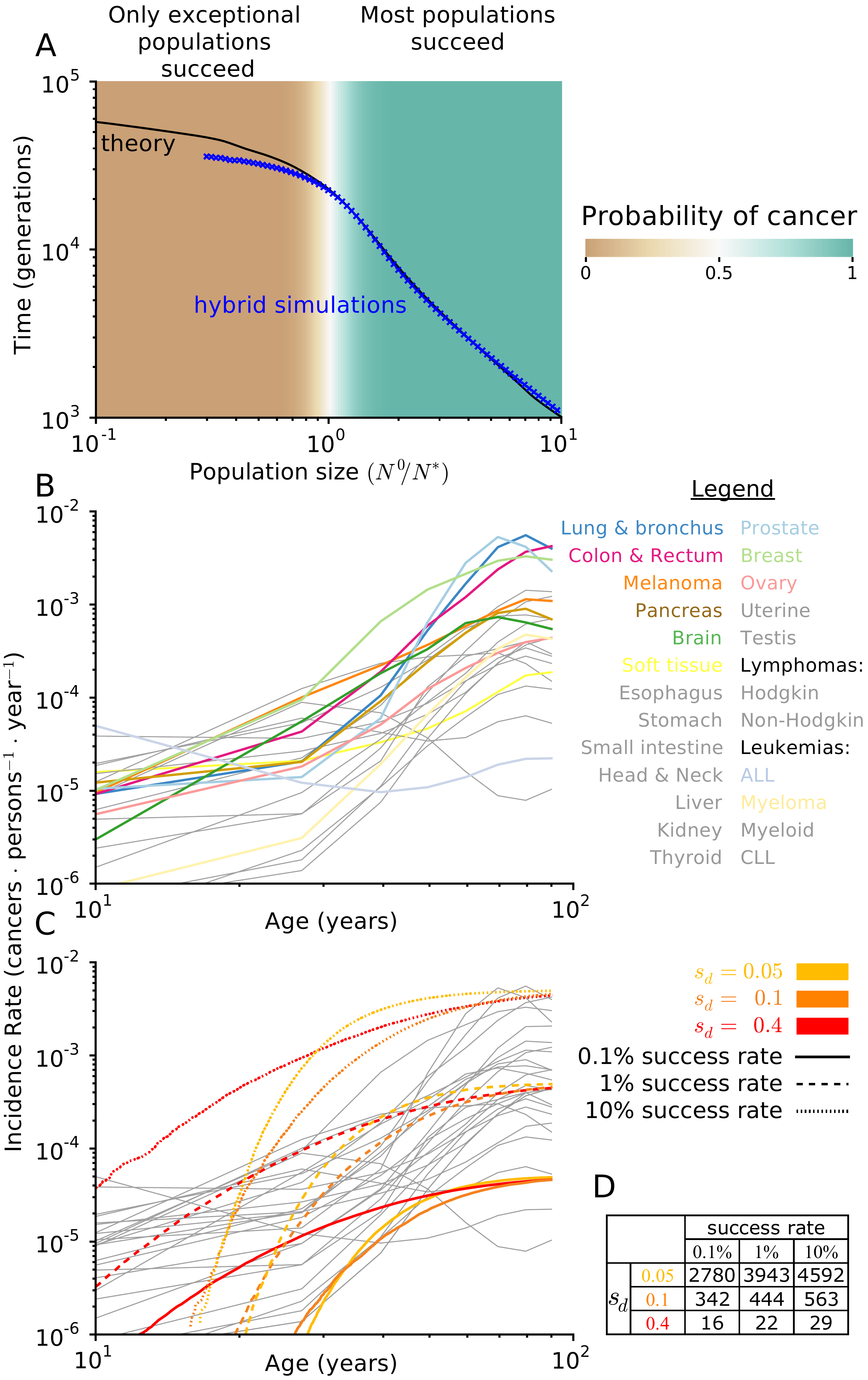}}
\caption{\textbf{Comparing waiting times to age-incidence curves}} 
\label{fig:incidence}
\end{center}
\end{figure}

The waiting times to cancer in this neutral-passenger model, has been previously shown to be a sum of exponentially-distributed waiting times \cite{ARMITAGE1954}, i.e. an Erlang or Gamma distribution, of the form: 
\begin{equation}\label{eq:power_law}
  \begin{array}{rr}
P(\tcancer| n_d=k)	=	& \text{Erlang}[\tcancer| n_d=k, r_d = 1] 			\\
						=	& \frac{r_d^k \tcancer^{k-1} e^{-r_d \tcancer}}{(k-1)!}	\\
			  			=	& \tcancer^{k-1} e^{-\tcancer } / (k-1)!				\\
				\propto 	& \Tcancer^{k-1} \mbox{ , when } \tcancer / k \ll 1 \\
  \end{array}
\end{equation}
Traditionally in this model, it is believed that very few precancerous population have enough time to progress, as lesion formation rates are much greater than cancer incidence rates. Hence, it is believed that age-incidence curves should be fit with only the beginning of this distribution: a power-law distribution presented in the last line is appropriate. We find that although this hypothesis explains age-incidence rates well at mid-age, it fails to explain the plateau in age-incidence rates seen at older ages in most cancer subtypes (\textbf{Fig. 2A, \ref{fig:incidence}}).  

In this model, the total number of passengers accumulated is a Poisson distribution, if the time of progression $\Tcancer$ is known:
\begin{equation}\label{eq:neutral_Poisson}
  \begin{array}{rr}
P(n_p| \tcancer, r_p) =	& \text{Poisson}[n_p| <n_p> = \Tcancer r_p] 	\\
			=	& e^{-<n_p>}<n_p>^{n_p}/n_p!			\\ 
  \end{array}
\end{equation}
Here, $<n_p> = \Tcancer r_p$ is the mean number of expected passengers. The distribution takes this form because each passenger accumulation event occurs with an exponentially-distributed waiting time; a Poisson distribution describes the sum of events with exponentially-distributed waiting times in a fixed time interval. Because we do not know when a new lesion arrises, we must convolute this distribution with our expected distribution of $\Tcancer$. 
 
The available time for cancer progression depends upon the length of a human life: $t_{\rm{human}}$. If $t_{\rm{human}} < \Tcancer$, then the precancerous population will be unobserved in age-incidence and genomics data because the person died of an alternate cause prior to malignancy. Although the actual distribution of human lifetimes is complicated, we can still make inferences about the validity of this model by considering its extremes. Consider two opposing extreme cases: (1) when $t_{\rm{human}} \gg \overline{\Tcancer}$, all lesions eventually progress and are sequenced (i.e. a human lifetime is much greater than the mean time to cancer); and (2) when $t_{\rm{human}} \ll \overline{\Tcancer}$, only a few exceptional lesions progress (i.e. the mean time to cancer is much shorter than a human lifetime). We find that this first extreme predicts a much broader and more positively skewed distribution in the number of passengers, than the second case (\textbf{Fig. \ref{fig:neutral_passengers}}). Nevertheless, it is still not wide enough, nor positively skewed enough, to explain the observed distribution of passengers in cancer under realistic parameters (\textbf{Fig. 2B, \ref{fig:neutral_passengers}, Table \ref{table:negative_binomial_fits}}). In contrast, our model predicts a broader and positively skewed distribution that captures observed passenger histograms well (\textbf{Fig. 2B}).

In the case where $t_{\rm{human}} \gg \overline{\Tcancer}$, accumulation of passengers follows a binomial process. Each accumulation event has probability $p=r_d/(r_d + r_p)$ of being a driver and probability $(1-p)$ of being a passenger. Because the population has infinite time to progress to cancer, the binomial process continues until $n_d=k$ drivers accumulate. A binomial process that continues until $k$ successes (i.e. drivers), will have a total number of failures (i.e. passengers) that samples a negative binomial distribution: 
\begin{equation}\label{eq:negative_binomial}
P(n_p| p, k) = \binom{n_p+k-1}{n_p} (1-p)^{n_p} p^k
\end{equation}
A negative binomial distribution with $p\ll1$ (i.e. passengers greatly outnumber drivers--as is the case in observed) reduces to a Poisson distribution. 

In the case where $t_{\rm{human}} \ll \overline{\Tcancer}$, the waiting time to cancer follows a power law distribution (Eq. \ref{eq:power_law}). This, convoluted with the distribution of passengers expected for a particular $\Tcancer$ (Eq. \ref{eq:neutral_Poisson}) yields the expected distribution of passengers for a cancer subtype:
\begin{equation}\label{eq:neutral_passenger_prediction}
  \begin{array}{rl}
P(n_p | k) =	& \int_0^{\tau_h}P(\tcancer|k)P(n_p|\tcancer, r_p) d\tcancer 	\\
		\approx		& \int_0^{\tau_h} \frac{\tcancer^{k-1} k}{\tau_h^k} 		
					\frac{e^{-\tau_p}\tau_p^{n_p}}{n_p!} d\tcancer 		 \\

		\approx		& 1/[*n_p!] \int_0^{\tau_h} e^{-\tau_p}
							\tcancer^{k-1} \tau_p^{n_p} d\tcancer 		\\		
		\approx		& k/[\tau_h^k n_p!r_p^k]  \int_{\tau_p=0}^{\tau_p=\tau_h r_p} e^{-\tau_p}
							\tau_p^{k-1 + n_p}			d\tau_p		\\	
		\approx		& \binom{n_p + k - 1}{n_p} k! {<n_p^{\rm{max}}>}^{-k}\  \gamma(k + n_p, <n_p>^{\rm{max}}) 

  \end{array}
\end{equation}
Where $\gamma(s, x)$ is the normalized incomplete gamma function defined previously (Eq. \ref{eq:Pcancer}). In the above derivation, we eliminated a parameter by considering the quantity: $<n_p>^{\rm{max}} = \tau_h r_p$, which corresponds to the mean number of passengers expected for a person who lives until the maximum allowable time, $\tau_h$. This solution is compared with the other extreme case in \textbf{Figure \ref{fig:neutral_passengers}}. Obviously, for both predicted passenger distributions (Eqs. \ref{eq:neutral_passenger_prediction} and \ref{eq:negative_binomial}) the total number of mutations $n_d + n_p$ is the expected number of passengers $P(n_p | k)$ \emph{plus} the number of drivers $k$, which is constant.







\bibliography{references}{}
\bibliographystyle{plain}



\clearpage
\noindent \textbf{A} Mean waiting time to cancer $\overline{\Tcancer } (x)$ decreases as initial population size $x = N^0/N^*$ increases. 
We solved for $\overline{\Tcancer } (x)$ from Eq. \ref{eq:diff} using two methods: (1) via stochastic `hybrid' Gillespie  simulations (Eq. \ref{eq:hybrid_gillespie}), and (2) via an analytical approximation (Eq. \ref{eq:progression_time}). Agreement between these two estimates suggests that our solution is accurate (see \textbf{Estimating the mean time to progression} for details).
The mean time to cancer $\overline{\Tcancer} (x)$ depends heavily on the probability of adaptation $\Pcancer(x)$, as $\overline{\Tcancer}$ is a conditioned on a population successfully progressing to cancer. Because $\Pcancer(x)$ has an inflection point at $x=1$, $\overline{\Tcancer}(x)$ behaves very differently when $x > 1$, than when $x < 1$. When $x>1$, $\overline{\Tcancer}$ is approximately the mean expected velocity of the population integrated over population size, as nearly all cancers succeed. However, when $x<1$, $\overline{\Tcancer}(x)$ is significantly shorter than would be expected from mean behavior. This is because most populations go extinct. Only the exceptional populations that progress to cancer are weighted in the mean of $\Tcancer(x)$; these exceptional populations happened to grow much faster than the average population. Hence, the increase in waiting time to cancer is sub-linear with respect to $x$ when $x<1$. 
Most importantly, these results suggest that the shape of our predicted age-incidence cures (below) will depend almost entirely on $s_d$ and not $x$, thereby simplifying the interpretation of data.  

\noindent \textbf{B} Incidence rate verses age for 25 most common cancers in the SEERs databases \cite{Howlader2013}. Nearly all cancers show incidence rates that rise rapidly at mid-life, but then plateau at old-age. Leukemias tend to have flatter curves, which may suggest that they need fewer drivers for carcinogenesis. Colorectal cancer is exceptional in that it does not plateau and, instead, exhibits a power-law relationship for all ages. Incidence curves are flatter at young ages because of patients with a genetic predisposition to cancer that expedites progression; neither our model or the traditional neutral-model of cancer progression attempt to explain these incidences.  

\noindent \textbf{C} The predicted age-incidence curves derived from simulations can explain observed age-incidence curves in most cancer subtypes well when proper parameters are chosen. The slope of predicted age-incidence curves is described by $s_d$: a larger $s_d$ causes the slope of age-incidence curves to decrease. In contrast, the location in the plateau of age-incidence curves is described by the success rate of cancer progression $P_{\infty}$ multiplied by the lesion formation rate $r$. While we do not know the precise value of these two quantities ($r$ and $P_{\infty}$) that determine the plateau level, we note that both parameters collectively introduce only \emph{one} free parameter in the description of age-incidence curves. Both parameters only alter the location of the plateau, not the shape of the distribution (as predicted in \textbf{A}). $r = 5$ in the predicted age-incidence curves plotted, however we expect this value to vary considerably between the cancer types. 
In the main text, we argue that $r$ is at least 10 $\mbox{lesions} \cdot \mbox{year}^{-1}$ in breast epithelial. This estimate was based on the assumption that $r = $(\# of mamospheres in breasts)(\# of stem cells per mamosphere)(\# of initiating mutations per cell per year) $= (10^7)(10)(10^{-7})$. Lower-bounds of literature-derived quantities were used in this estimate [,,]. Moreover, the number of lesions observed in normal breasts corroborates this estimate \cite{Page1985, Wellings1975}. Thus, we eliminate the possibility that age-incidence curves can be explained by models which assume that most lesions eventually progress to cancer. 

\noindent \textbf{D} The actual initial population size $(N^0)$ necessary to obtain various success rates of cancer progression from various $s_d$ utilized in \textbf{C}. Values were obtained by iteratively simulating various initial sizes until converging to the desired success rate.

\clearpage

\begin{figure}[ht]
\begin{center}
\centerline{\includegraphics[width=.5\textwidth]{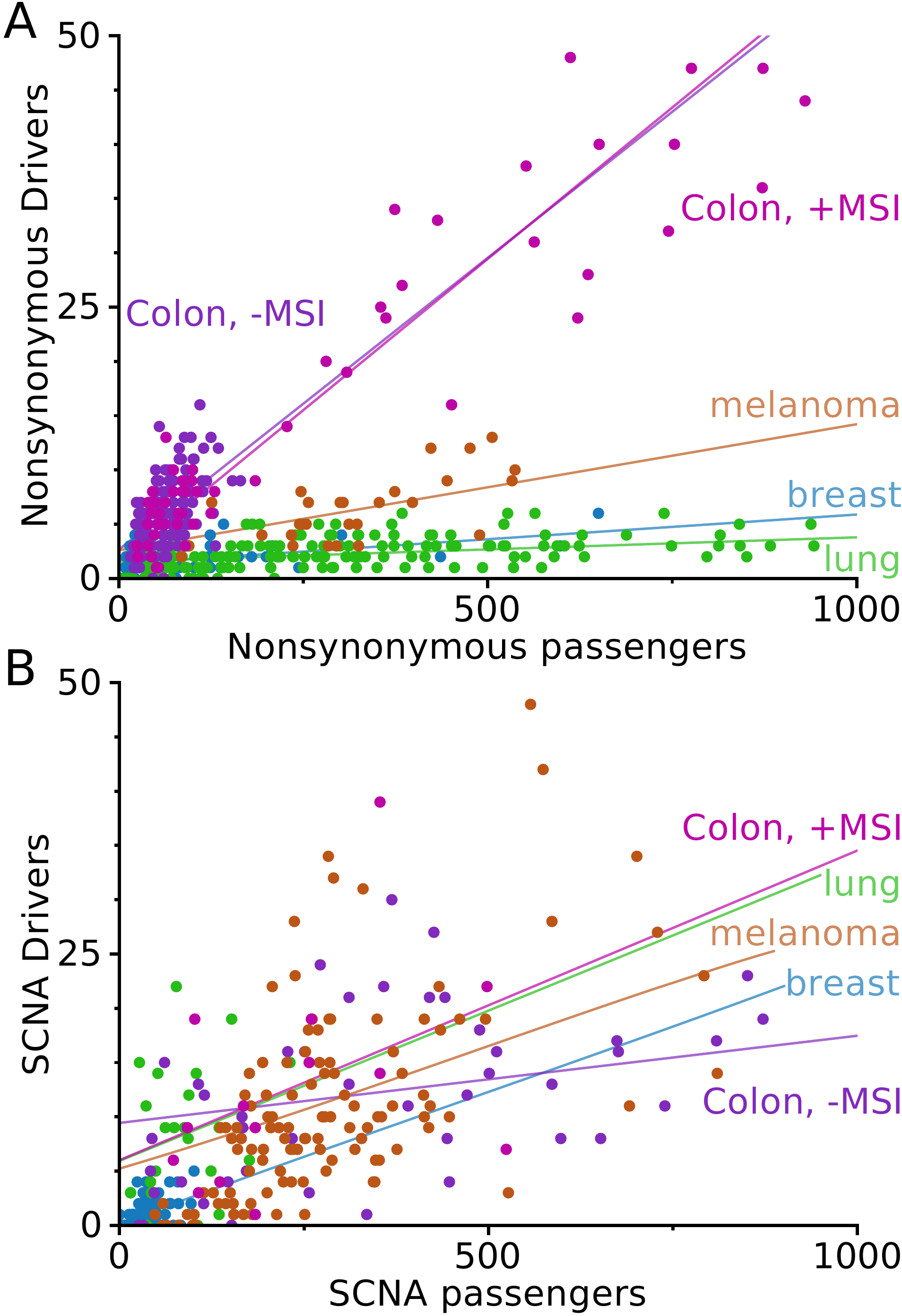}}
\caption{\textbf{Comparison of drivers and passengers in Somatic Nonsynonymous Mutations (SNMs) and Somatic Copy Number Alterations (SCNAs).}
\textbf{A} A positive linear relationship is observed between driver and passenger SNMs in all cancer subtypes studied here. This suggests that additional SNM passengers are being counterbalanced by additional drivers, and is consistent with our conclusions in the main text. Slope and y-intercept of the best-fit lines can be found in \textbf{Table \ref{table:correlations}}. 
\textbf{B} This positive linear relationship is also observed in SCNAs. The similar slopes and y-intercepts of SCNAs to SNMs supports our assumption that SCNAs and SNMs can be aggregated in analysis and modeling.}
\label{fig:SNM_SCNA_nd_np}
\end{center}
\end{figure}

\clearpage

\begin{figure}[ht]
\begin{center}
\centerline{\includegraphics[width=1.0\textwidth]{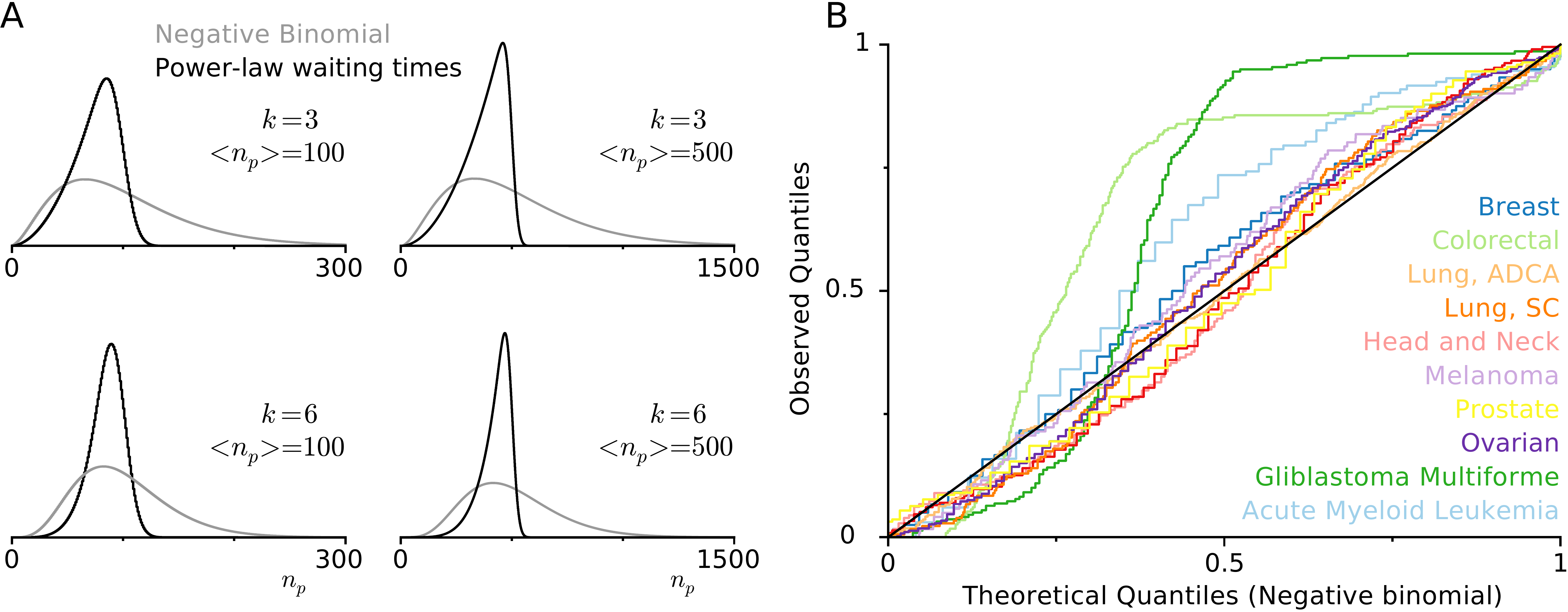}}
\caption{\textbf{Neutral-passenger model of cancer progression explains distribution of mutation totals in cancers only with unrealistic parameters.}}
\label{fig:neutral_passengers}
\end{center}
\end{figure}
\noindent \textbf{A} We estimated the expected distribution of total mutations in sequenced tumors from a traditional model of cancer progression with neutral passengers (see \textbf{A traditional model of cancer progression with drivers and neutral passengers.} for details). The expected distribution depends upon the time allowed for progression. If the time for progression is much shorter than the mean, we expect power-law waiting times to cancer and a distribution of passengers that follows Eq. \ref{eq:neutral_passenger_prediction} (\textbf{Black lines}). If the time to progression is much longer than the mean, then we expect a Negative Binomial distribution (Eq. \ref{eq:negative_binomial}, \textbf{Gray lines}). We compared these distributions for various quantities of drivers necessary for carcinogenesis ($k=3,\ 5$) and various quantities of the mean number of passengers ($<n_p^{\rm{max}}> = 100,\ 500$). Because these distributions differed and we found no compelling reason to chose one over the other, we selected the distribution that was \emph{a priori} more likely to fit the observed distributions of total mutations; we wanted to give the neutral-passenger model of cancer progression every opportunity to succeed. 
The second scenario (gray lines), where cancers have ample time to progress predicts a distribution with more variance and positive skew.  Large variance and positive skew are observed in the true distribution of passengers, therefore we used this distribution.
\textbf{B} We investigated the mutation totals of the 11 cancer subtypes that TCGA has sequenced in over 100 tumors \cite{Lawrence2013}. 
The distribution of mutation totals in these subtypes is compared in a Quantile-Quantile plot to the best-fitting Negative Binomial distribution for each subtype. 
Deviations from the black line indicate regions of the observed distribution that are poorly explained by the theoretical distribution. 
While this theoretical distribution explains observed distributions well (excluding Colorectal cancers and Gliablastoma Multiforme), the Maximum-Likelihood Estimators (MLEs) used to fit this distribution suggest that only 1-2 drivers are necessary for progression (\textbf{Table \ref{table:negative_binomial_fits}})---quantities that are inconstant with known biology and age-incidence curves.

\clearpage

\begin{figure}[ht]
\begin{center}
\centerline{\includegraphics[width=\textwidth]{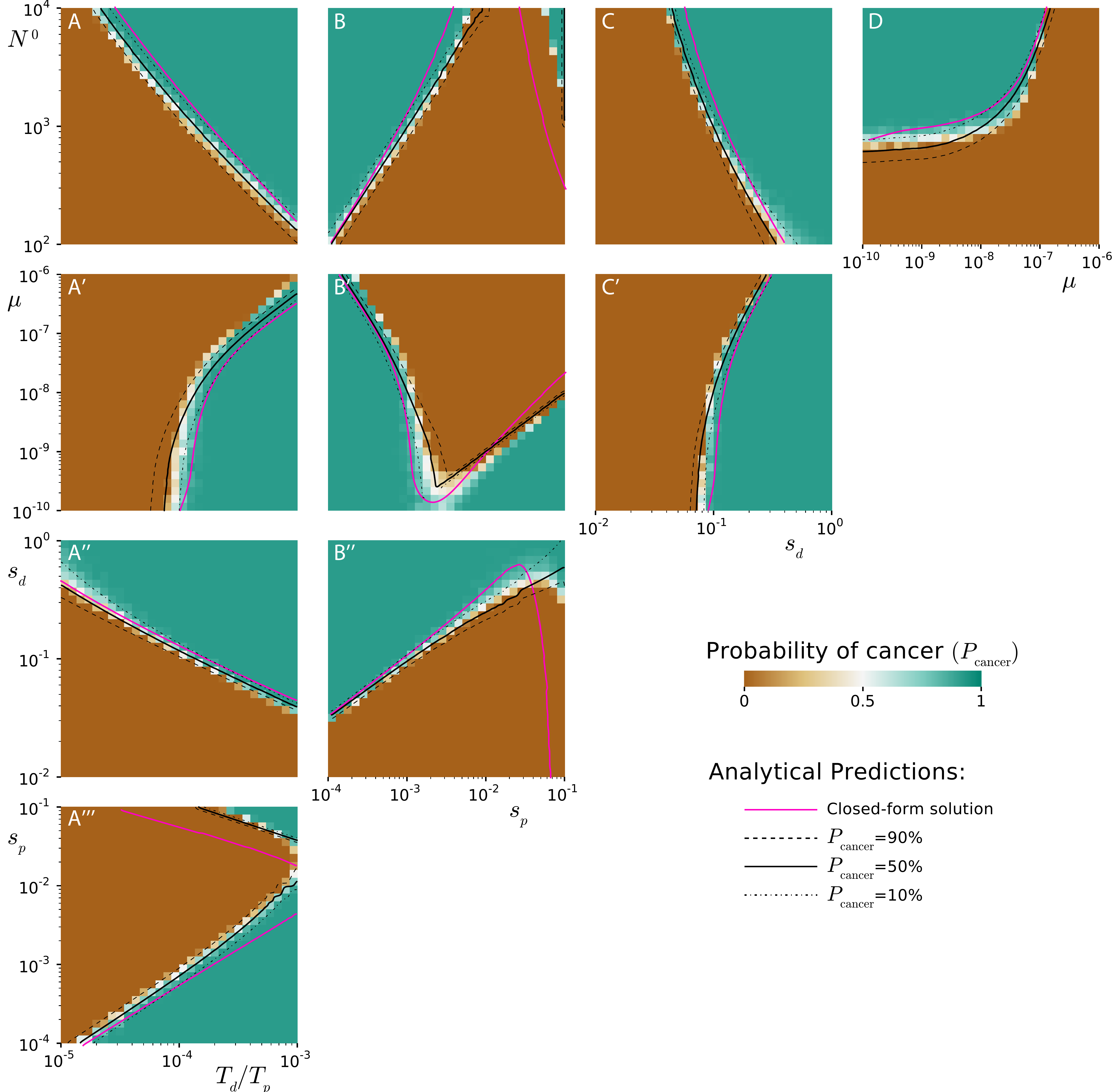}}
\caption{\textbf{Analytical framework predicts probability of cancer across parameter space.}}
\label{fig:phase_space}
\end{center}
\end{figure}

\noindent The probability of progression, determined from the outcome of 3,000 simulations propagated until extinction or rapid growth, across the parameter range of our model. In simulations, we observe parameters where progression occurs, fails, and is rare. A sophisticated analytical framework incorporating selection against passengers, hitchhiking of passengers onto driver mutations, and stochasticity predicts phase patterns well (black lines). 
This sophisticated analytical model uses two solutions for Muller's Ratchet in various parameter regimes (see \textbf{Selection against passengers.}), an estimate of the quantity of hitchhiking passengers (see \textbf{Effects of passengers on driver fixation}), and a stochastic differential equation to estimate probabilities of progression (see \textbf{Estimating the probability of cancer}). 
A simplified framework, which offers a closed-form solution, is possible and works reasonably well (magenta). This solution differs from the more precise solution in two ways: (1) a novel estimate of Muller's ratchet is used (Eq. \ref{eq:kirill_ratchet}), and (2) hitchhikers that accumulate after the new driver clone arises $\delta p_{\rm{S}}$ are neglected. 
\textbf{A-A$'''$} $\Pcancer$ increases as the relative target size of drivers $T_d$ verses passengers $T_p$ increases, for all parameters. 
\textbf{B-B$'''$, A$'''$} $\Pcancer$ exhibits a local minimum with respect to the selection against passengers $s_p$. When the selection against passengers is very weak, passengers are effectively neutral. When the selection against passengers is too strong, natural selection weeds out passengers and they never accumulate. Deleterious passengers are most effective at preventing cancer when at moderate fitness effects. This local minimum suggests that there may be two types of cancers: those that exist in an environment/genetic context where passengers are weak and buffered by, for example, an activated UPR; and those that succeed by exacerbating passengers' deleterious effects, perhaps by minimizing the mutation rate. 
\textbf{C, C$'$, A$''$, B$''$} The probability of cancer always increases with $s_d$. This parameter has a profound affect on cancer progression, as can be seen by how little the boundary between success and failure appears to be almost independent of the other parameters. Thus, this parameter is easy to estimate from epidemiological patterns. 
\textbf{D, A$'$} An increasing mutation rate affects the probability of cancer very little at first; however, once it exceeds a critical value ($\mu^*$), the probability of cancer drops precipitously.

\clearpage

\begin{figure}[ht]
\begin{center}
\centerline{\includegraphics[width=\textwidth]{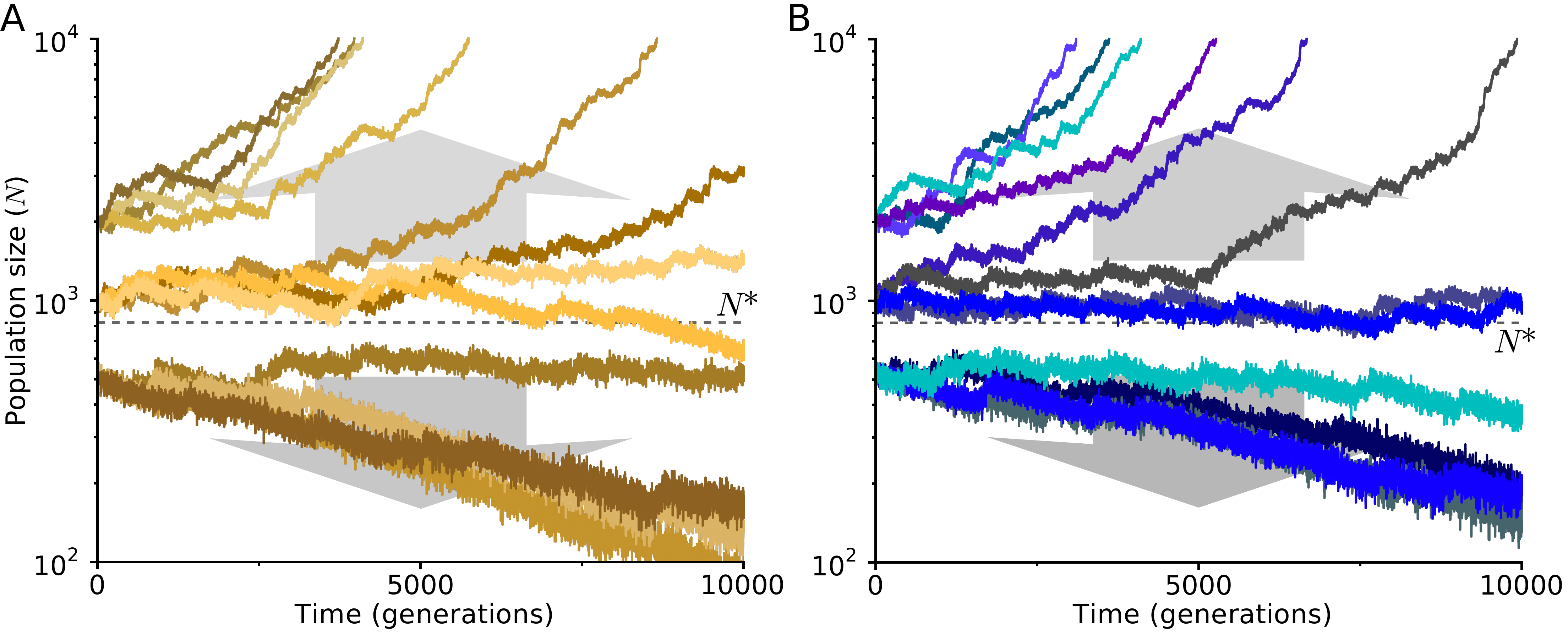}}
\caption{\textbf{Simulations exhibit path independence}.
\textbf{A} The 12 trajectories from Figure 1A in the main text, initiated at $N^0 = \{500, 1000, 2000\}$. 
\textbf{B} An additional 12 trajectories, initiated at various $N^0$, but plotted once they cross $N=\{500, 1000, 2000\}$. Populations that crossed $N = 500$ and $N = 2000$ were initiated at $N^0 = 1000$, while populations that crossed $N = 1000$ were initiated at $N^0=500$.
This comparison demonstrates that populations beginning at different initial sizes $N^0$ will behave similarly, if they have the same current size. Thus, populations evolve independent of their past history and the entire state of a trajectory is defined by $x = N/N^0$.}
\label{fig:path_independence}
\end{center}
\end{figure}

\clearpage

\begin{figure}[ht]
\begin{center}
\centerline{\includegraphics[width=\textwidth]{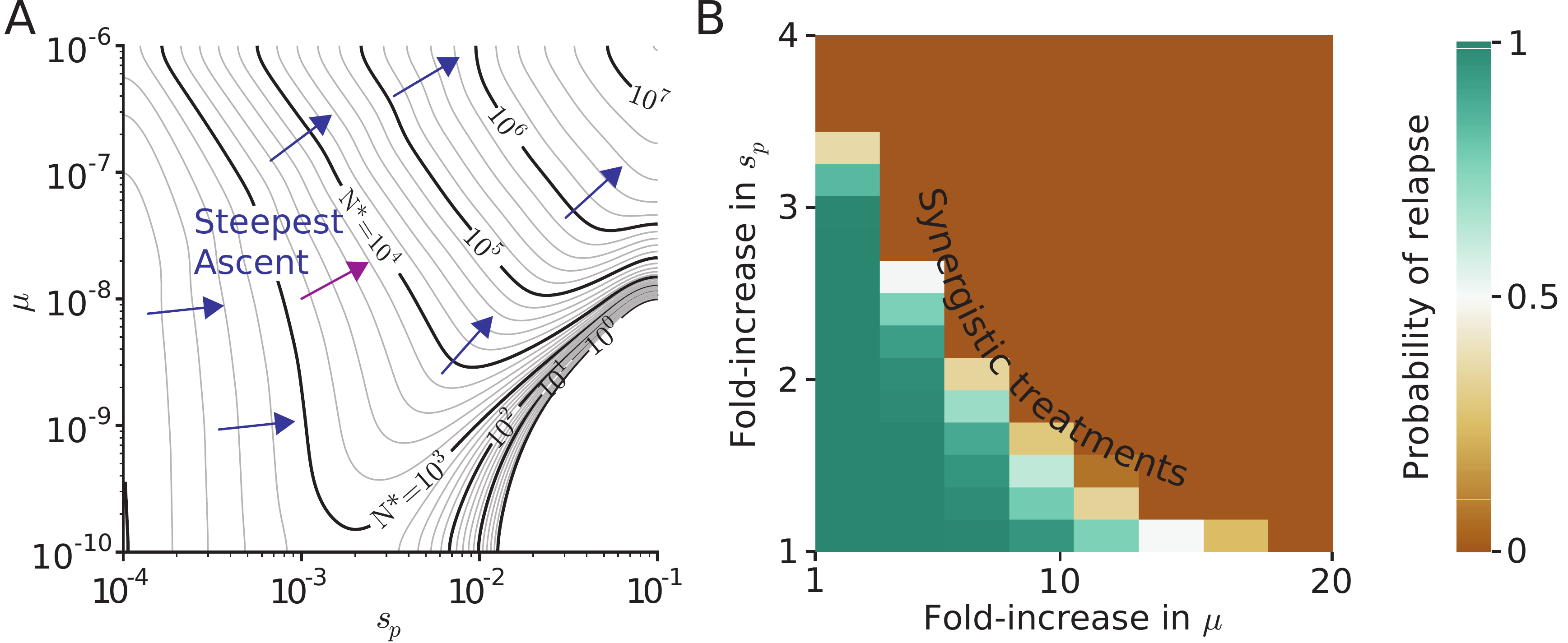}}
\caption{\textbf{Combination treatments that increase mutation rate and selection against passengers work best.} 
\textbf{A} Using the analytical theory describe in \textbf{Fig. \ref{fig:phase_space}}, we can plot the critical population size $N^*$ across evolutionary parameters as a contour plot. Optimal therapy would be drug combinations that increase $N^*$ most dramatically, which would be the gradient of steepest ascent (blue lines). From this 3-Dimensional perspective the interplay between $\mu$ and $s_p$ is evident. At low mutation rates, only weak passengers with low $s_p$ can fixate. Thus, treating cancers with low mutation rates by increasing $s_p$ may be ineffective. At high mutation rates, all passengers fixate and increasing $s_p$ increases $N^*$. At intermediate mutation rates, the most effective way to increase $N^*$ would be to moderately increase both the mutation rate and $s_p$.  
\textbf{B} Via simulations, we tested our prediction that the gradient of steepest ascent is optimal for the magenta gradient line in \textbf{A}. 50 cancers with $\mu=10^{-8}, s_p=0.001$ grown to $10^6$ cells were treated with combinations of mutagenic and $s_p$ increasing therapy. Indeed, moderate increases in both parameters were more effective than would be expected from the lone treatments, thus confirming our prediction. These results suggest that combinatorial therapies may be most effective at treating cancer and that evolutionary modeling can guide clinical decisions. }
\label{fig:synergistic_treatment}
\end{center}
\end{figure}


\clearpage

\begin{table}
\begin{center}
\caption{\textbf{Evolutionary parameters explored in this study}} 
\begin{tabular}{lccc@{-}cc}
\toprule
Parameter 		& 	Symbol 	& 	Estimate 	& 	\multicolumn{2}{c}{Range} 					& 	Citation 									\\
\midrule
Mutation rate 		&	$\mu$	&	$10^{-8}$		&	$10^{-10}$ & $10^{-7}$			&	\cite{Loeb2008} 							\\
Driver loci 			& 	$T_d$ 	& 	$700$			&	70 & 7,000						&	\cite{Futreal2004,Sjoblom2006,Beerenwinkel2007}	\\
Passenger loci		&	$T_p$	&	$5\cdot10^{6}$	&	$5\cdot 10^{5}$ & $5\cdot10^{7}$	&	\cite{Beckman2005,Boyko2008}					\\
Driver strength		&	$s_d$	&	$0.1$			&	0.001 & 1						&	\cite{Beerenwinkel2007,Beroukhim2010}			\\
Passenger strength	&	$s_p$ 	&	$0.001$			&	$10^{-4}$ & $10^{-1}$ 				&	\cite{Geiler-Samerotte2011}					\\
Initial population size	&	$N^0$ 	&	$1000$*			&	100 & 10,000						&	\cite{Cole2010}								\\
\bottomrule
\multicolumn{6}{l}{*Estimated from labeled populations in mice colonic crypts 2 weeks}				\\
\multicolumn{6}{l}{$\ $after an initiating \emph{APC} deletion was induced.}						\\
\end{tabular}
\end{center}
We explored our evolutionary model incorporating driver and passenger mutations across a broad range of parameters. The ranges were motivated by literature estimates that we discussed previously \cite{McFarland2013}. Note that in simulations $\mu_d = \mu T_d$ and $\mu_p = \mu T_p$), hence we can explore the entire phase space by only altering $\mu$ and $T_d/T_p$; altering all three parameters would be redundant. In \textbf{Figure 2} we compare our model to epidemiological and genomic data and find that the best-fitting parameters agree well with these prior published estimates.
\label{table:parameters}
\end{table}

\clearpage

\begin{table}
\begin{center}
\caption{\textbf{Average number of driver and passenger mutations by tumor type}} 
\begin{tabular}{lrrrr}
\toprule 
		& NSM 		& NSM 			& SCNA 		& SCNA 			\\
Cancer 	& Drivers 	& Passengers 	& Drivers 	& Passengers 	\\
\midrule
breast & 1.7 & 70.8 & 1.0 & 34.6 \\
lung & 2.3 & 348.6 & 8.4 & 89.5 \\
colon, MIN$^-$ & 8.8 & 114.0 & 14.1 & 583.5 \\
colon, MIN$^+$ & 28.8 & 489.0 & 12.7 & 235.1 \\
melanoma & 7.0 & 379.6 & 12.6 & 324.7 \\
\hline
all & 9.1 & 272.8 & 8.8 & 258.9 \\
Max & 28.8 & 489.0 & 14.1 & 583.5 \\
Min & 1.7 & 70.8 & 1.0 & 34.6 \\
\bottomrule
\end{tabular}
\end{center}
The total number of identified Somatic Nonsynonymous Mutations (SNMs) and Somatic Copy Number Alterations (SCNAs) for various tumor-normal paired sequences from various tissues of origin: 100 breast \cite{Stephens2012}, 183 lung \cite{Ding2008}, 159 Colon without Micro-satellite INstability (MIN$^-$), 64 Colon with Micro-Satellite Instability (MIN$^+$) \cite{Cancer2012}, and 121 melanomas \cite{Berger2012}. 
\label{table:alterations_summary}
\end{table}

\newcommand{\gray}[1] {
  \textcolor[rgb]{0.4,0.4,0.4}{#1}}

\clearpage

\makeatletter
\newcommand{\thickhline}{%
    \noalign {\ifnum 0=`}\fi \hrule height 2pt
    \futurelet \reserved@a \@xhline
}
\newcolumntype{"}{@{\hskip\tabcolsep\vrule width 1pt\hskip\tabcolsep}}
\makeatother

\begin{table}
\begin{center}
\caption{\textbf{Linear relationship between drivers and passengers cannot be explained by other tumor properties}} 
\begin{tabular}{lrcrrrr}
\toprule
Cancer	&	Pearson's $r$	&	$p$-value*	&	 $N\dagger$ 	&Spearman's $\rho$	&	slope ($s_p/s_d$)		&	y-intercept	\\
\midrule
    \multicolumn{7}{c}{\textbf{Drivers verses Passengers}} \\
\hline
    breast & 0.423 & $\bf{<10^{-4}}$ & 100   & 0.413 & 0.006 & 2.02 \\
    lung  & 0.368 & 0.08  & 24    & 0.998 & 0.005 & 8.63 \\
    colon, MIN$^-$ & 0.624 & $\bf{<10^{-4}}$ & 49    & 0.985 & 0.009 & 17.11 \\
    colon, MIN$^+$ & 0.916 & $\bf{<10^{-5}}$ & 14    & 0.999 & 0.047 & 6.50 \\
    melanoma & 0.749 & $\bf{<10^{-5}}$ & 29    & 0.995 & 0.015 & 3.69 \\
    All   & 0.937 & $\bf{<10^{-99}}$ & 217   & 0.992 & 0.042 & \gray{-3.81} \\
\hline
    \multicolumn{7}{c}{\textbf{SNM drivers versus SNM passengers}} \\
\hline
    breast & 0.390 & $\bf{<10^{-4}}$ & 100   & 0.178 & 0.005 & 1.34 \\
    lung  & 0.587 & $\bf{<10^{-17}}$ & 183   & 0.579 & 0.002 & 1.56 \\
    colon, MIN$^-$ & 0.990 & $\bf{<10^{-134}}$ & 159   & 0.569 & 0.054 & 2.65 \\
    colon, MIN$^+$ & 0.994 & $\bf{<10^{-60}}$ & 64    & 0.918 & 0.056 & 1.56 \\
    melanoma & 0.878 & $\bf{<10^{-9}}$ & 29    & 0.974 & 0.012 & 2.59 \\
    All   & 0.924 & $\bf{<10^{-223}}$ & 536   & 0.592 & 0.050 & \gray{-4.45} \\
\hline
    \multicolumn{7}{c}{\textbf{SCNA drivers verses SCNA passengers}} \\
\hline
    breast & 0.443 & $\bf{<10^{-5}}$ & 100   & 0.433 & 0.024 & 0.17 \\
    lung  & 0.253 & 0.23  & 24    & 0.998 & 0.028 & 5.94 \\
    colon, MIN$^-$ & 0.770 & $\bf{<10^{-9}}$ & 49    & 0.984 & 0.008 & 9.44 \\
    colon, MIN$^+$ & 0.424 & 0.13  & 14    & 0.994 & 0.029 & 6.01 \\
    melanoma & 0.559 & $\bf{<10^{-10}}$ & 121   & 0.663 & 0.023 & 5.23 \\
    All   & 0.573 & $\bf{<10^{-27}}$ & 309   & 0.962 & 0.012 & 5.76 \\
\thickhline
\multicolumn{7}{c}{\textbf{SNMs versus SCNAs}}								\\
\hline
breast	&	0.052		&	0.61		&	100	&	0.149		&	0.237		&	64	\\
lung		&	0.169		&	0.43		&	24	&	\gray{-0.548}	&	1.268		&	334	\\
colon, MIN$^-$ &	\gray{-0.080}	&	0.58		&	49	&	\gray{-0.068}	&	\gray{-0.021}	&	137	\\
colon, MIN$^+$ &	\gray{-0.265}	&	0.36		&	14	&	0.045		&	\gray{-0.981}	&	838	\\
melanoma	&	\gray{-0.114}	&	0.56		&	29	&	0.176		&	\gray{-0.183}	&	431	\\
All		&	0.331	&	$\bf{<10^{-6}}$&	217	&	\gray{-0.089}	&	0.631		&	133	\\
\hline
    \multicolumn{7}{c}{\textbf{Drivers verses Pathological Grade}} \\
\hline
    breast & 0.163 & 0.10  & 100   & 0.113 & 0.067 & 2.25 \\
    lung  & \gray{-0.048} & 0.83  & 22    & 0.024 & 0.006 & 2.13 \\
    colon, MIN$^-$ & \gray{-0.187} & 0.20  & 48    & 0.072 & 0.012 & 2.67 \\
    colon, MIN$^+$ & \gray{-0.121} & 0.68  & 14    & \gray{-0.338} & 0.004 & 3.09 \\
    melanoma & 0.221 & 0.35  & 20    & 0.120 & 0.025 & 1.83 \\
    All   & 0.018 & 0.80  & 204   & 0.054 & 0.001 & 2.37 \\
\hline
    \multicolumn{7}{c}{\textbf{SNMs versus Pathological Grade}} \\
\hline
    breast & 0.217 & \textbf{0.03} & 100   & 0.444 & 0.001 & 2.33 \\
    lung  & 0.193 & \textbf{0.02} & 158   & 0.235 & 0.000 & 1.79 \\
    colon, MIN$^-$ & \gray{-0.084} & 0.30  & 156   & 0.039 & 0.000 & 2.51 \\
    colon, MIN$^+$ & \gray{-0.045} & 0.73  & 63    & \gray{-0.023} & 0.000 & 2.50 \\
    melanoma & 0.119 & 0.62  & 20    & 0.114 & 0.000 & 2.06 \\
    All   & \gray{-0.012} & 0.80  & 497   & 0.147 & 0.000 & 2.28 \\
\hline
    \multicolumn{7}{c}{\textbf{SCNAs versus Pathological Grade}} \\
\hline
    breast & 0.248 & \textbf{0.01} & 100   & 0.235 & 0.007 & 2.18 \\
    lung  & 0.170 & 0.45  & 22    & 0.026 & 0.002 & 1.85 \\
    colon, MIN$^-$ & \gray{-0.166} & 0.26  & 48    & 0.081 & \gray{0.000} & 2.48 \\
    colon, MIN$^+$ & \gray{-0.054} & 0.85  & 14    & \gray{-0.333} & \gray{0.000} & 2.99 \\
    melanoma & 0.109 & 0.31  & 88    & \gray{-0.253} & 0.000 & 2.06 \\
    All   & \gray{-0.067} & 0.27  & 272   & \gray{-0.112} & \gray{0.000} & 2.38 \\
\bottomrule
\multicolumn{7}{l}{*Statistically significant ($p<0.05$) relationship are in bold}	\\
\multicolumn{7}{l}{$\dagger$ Number of samples compared}	\\
\end{tabular}
\label{table:correlations}
\end{center}
\end{table}
\clearpage

\noindent Negative values are in gray. We observe a linear relationship between drivers and passengers that was predicted by our model. Above the thick black line are relationships that appear to robustly covary, while the bottom half contains relationships that we believe are insignificant. 
In our model, driver's and passenger's linear relationship results from their competing effect: additional deleterious passengers must be overcome by additional drivers. However, this relationship could conceivably be explained by alternate factors; in particular, we were concerned that the mutation type, mutation rate, or aggressiveness of the tumor could also explain the observed relationship. The data above suggest that these possibilities are unlikely, thus supporting our conclusion that drivers compete with passengers. Our rational for the other competing hypotheses and why we reject them: 
(a) SCNAs and SNMs might have drastically different effects on cancer progression and undermine our model. The slope and y-intercept between drivers and passengers is \emph{approximately} equal in SCNAs and SNMs, suggesting the relative fitness effects of these mutations is similar. 
(b) Some cancers might progress via CIN, while others progress via an elevated point mutation rate. If so, a negative correlation between SCNAs and SNMs within tumor subtypes would be expected, which has been observed previously in a pan-cancer study \cite{Ciriello2013} and within the aggregate colorectal dataset. However, this does not appear to be so in other tumor types, nor in colorectal cancer after segregation according to MIN phenotype. Thus, the observed patterns are not explicable by varying mutational mechanisms. 
(c) The relationship between drivers and passengers might be a result of variation in mutation rate. Variation in the mutation rate should only alter the waiting time to cancer in the neutral-passenger model, and not alter mutation totals. Nevertheless, if variation in the mutation rate could explain the correlation between drivers and passengers, then stratifying tumors by their mutation rate should reduce the correlation. Because the relationship between drivers and passengers is persistent and strong within the MIN$^+$ and MIN$^-$subtypes---expected to have and not-have a mutator phenotype---we reject this hypothesis. 
(d) Tumors with more drivers and passengers might simply be more evolutionarily advance. Suppose some cancers are detected and sequenced later than others. These late cancers would not only possess additional drivers, but also additional passengers, even if passengers were neutral; thus, retaining the correlation between drivers and passengers. However, late-detected tumors with additional drivers should also be more advanced and more aggressive. We find that a tumor's pathological grade is uncorrelated with the number of drivers, refuting this possibility. Pathological grade was quantified by converting roman numerals into a linear scale (i.e. A Stage IV tumor corresponds to an aggressiveness of 4). Many tumors had intermediate grades that were given corresponding fractional values (e.g. a Stage IIIa tumor was translated into a 3.0, a Stage IIIb was given 3.3, and a Stage IIIc was given 3.7). Because this quantification of tumor grade may distort the scale of aggressiveness, Spearman's Rank correlations are provided.
For completeness, we have also show the relationship between SNMs and Pathological Grade and SCNAs and Pathological Grade.

\clearpage

\begin{table}
\begin{center}
\caption{\textbf{Maximum-Likelihood Estimates (MLE) of $\bf{n_d}$ using neutral-passenger model of cancer progression}}
\begin{tabular}{lrccl}
\toprule
 & &Total mutations& \multicolumn{2}{l}{MLE of}  \\ 
Tissue&$N$&(mean $\pm$ STD) & $n_d$ & $p$* \\
\midrule
Rhabdoid tumor			& 20  & 7.8 $\pm$ 8.93               & 1         & 0.13     \\
Thyroid					& 52  & 11 $\pm$ 6.13               & 2         & 0.18     \\
Medulloblastoma			& 26  & 11.1 $\pm$ 4.9               & 4         & 0.36     \\
Ewing Sarcoma			& 20  & 12.6 $\pm$ 10.8               & 2         & 0.16     \\
Carcinoid					& 23  & 19.3 $\pm$ 9.09               & 3         & 0.16     \\
Neuroblastoma			& 80  & 21.2 $\pm$ 27                 & 1         & 0.047    \\
\textbf{Acute myeloid leukemia}		& \textbf{132} & \textbf{24.2 $\pm$ 57.6}               & \textbf{1}         & \textbf{0.041}    \\
Chronic lymphocytic leukemia & 91 & 25.3 $\pm$ 15.3               & 3         & 0.12     \\
\textbf{Prostate}			& \textbf{221} & \textbf{25.6 $\pm$ 26.2}   & \textbf{2}         & \textbf{0.078}    \\
Pancreas                      & 12  & 28.8 $\pm$ 18.4               & 1         & 0.035    \\
Low-grade glioma              & 57  & 29.6 $\pm$ 14.4               & 3         & 0.1      \\
\textbf{Breast}                        & \textbf{120} & \textbf{39.2 $\pm$ 39.8}               & \textbf{2}         & \textbf{0.051}    \\
Multiple myeloma              & 63  & 48.1 $\pm$ 41.4               & 2         & 0.042    \\
\textbf{Kidney clear cell}             & \textbf{214} & \textbf{50.4 $\pm$ 41.4}               & \textbf{3}         & \textbf{0.06}     \\
Kidney papillary cell         & 11  & 52.1 $\pm$ 19.4               & 4         & 0.077    \\
\textbf{Ovarian}                       & \textbf{385} & \textbf{64 $\pm$ 73.8}               & \textbf{2}         & \textbf{0.031}    \\
Diffuse large B-cell lymphoma & 49  & 114 $\pm$ 73.3               & 2         & 0.018    \\
\textbf{Head and neck}                 & \textbf{178} & \textbf{127 $\pm$ 129}                & \textbf{1}         & \textbf{0.0079}   \\
\textbf{Glioblastoma multiforme}       & \textbf{219} & \textbf{134 $\pm$ 949}                & \textbf{1}         & \textbf{0.0075}   \\
Esophageal adenocarcinoma     & 76  & 165 $\pm$ 236                & 2         & 0.012    \\
Cervical                      & 20  & 187 $\pm$ 292                & 1         & 0.0054   \\
Bladder                       & 35  & 237 $\pm$ 298                & 2         & 0.0084   \\
\textbf{Lung adenocarcinoma}           & \textbf{333} & \textbf{282 $\pm$ 294}                & \textbf{1}         & \textbf{0.0036}   \\
\textbf{Colorectal}                    & \textbf{230} & \textbf{313 $\pm$ 976}                & \textbf{1}         & \textbf{0.0032}   \\
\textbf{Lung squamous cell}  & \textbf{178} & \textbf{316 $\pm$ 307}                & \textbf{2}         & \textbf{0.0063}   \\
Stomach                       & 88  & 441 $\pm$ 673                & 1         & 0.0023   \\
\textbf{Melanoma}                      & \textbf{121} & \textbf{680 $\pm$ 934}                & \textbf{1}         & \textbf{0.0015}  \\
\bottomrule
\multicolumn{5}{l}{*$p=\frac{r_d}{r_d + r_p}$ i.e. the probability of an accrued mutation being a driver.} \\
\end{tabular}
\end{center}
A classical model of cancer progression with neutral passengers predicts that the distribution of passengers will follow a Negative Binomial distribution (see \textbf{Fig. \ref{fig:neutral_passengers}} and \textbf{A traditional model of cancer progression with drivers and neutral passengers.} for details). Tissues are ordered according to mean number of mutations and tissues with >100 samples are in \textbf{bold}. The MLE fits to observed distributions predict that only 1-3 drivers are needed for carcinogenesis. This is far fewer than expected from age-incidence curves, known biology, and driver classification algorithms. Therefore, we believe that the neutral-passenger model of cancer progression can not explain the observed spectrum of mutation totals.
\label{table:negative_binomial_fits}
\end{table}

\clearpage

\end{document}